\documentclass{jfm}

\usepackage{graphicx}
\usepackage{newtxtext}
\usepackage{newtxmath}
\usepackage{natbib}
\usepackage{hyperref}
\hypersetup{
    colorlinks = true,
    urlcolor   = blue,
    citecolor  = black,
}

\newcommand{\RomanNumeralCaps}[1]
\linenumbers

\usepackage{xcolor}
\usepackage{subcaption}

\usepackage{tikz}
\usetikzlibrary{arrows.meta,calc,decorations.pathmorphing,decorations.markings,positioning}
\usetikzlibrary{calc}

\usepackage{array}
\usepackage{multirow}

\usepackage{bm}
\usepackage{tabularx}
\usepackage{siunitx}
\usepackage{xfrac}
\usepackage{soul}

\input{linestyles.tex}

\newcommand{\rebulk} {\Rey_\mathrm{b}}
\newcommand{\retau} {\Rey_\tau}
\newcommand{\ubulk} {u_\mathrm{b}}
\newcommand{\utau} {u_\tau}
\newcommand{\tauwall} {\tau_{\mathrm{w}}}
\newcommand{\rms} {\mathrm{rms}}
\newcommand{\deltasil}{\delta_{\mathrm{SIL}}}
\newcommand{\deltanvd}{\delta_{\mathrm{NVD}}}
\newcommand{\deltavisc}{\delta_{\mathrm{V}}}
\newcommand{\deltak}{\delta_{\mathrm{K}}}
\newcommand{\lenvisc}{\ell_{\mathrm{V}}}
\newcommand{\lenk}{\ell_{\mathrm{K}}}
\title{How is the free surface influence transported in turbulent open channel flows?}

\author{
Yoshiyuki Sakai\aff{1}
\corresp{\email{yoshiyuki.sakai@tum.de}}
\and
Christian Bauer\aff{2}
}

\affiliation{
\aff{1}
Technical University of Munich, TUM School of Engineering and Design, TUM Department of Civil, Geo and Environmental Engineering, Arcisstr. 21, 80333 Munich, Germany
\aff{2}
German Aerospace Center (DLR), Institute of Aerodynamics and Flow Technology, Bunsenstr. 10, 37073 G\"ottingen, Germany
}
\begin{document}
\maketitle

We investigate how the influence of a free surface is transported in turbulent open channel flow by analysing matched open- and closed-channel direct numerical simulations up to $\retau \approx 900$ in a domain large enough to accommodate very-large-scale motions (VLSMs). 
The turbulent kinetic energy (TKE) budget shows that the surface influence is communicated primarily through transport terms. 
Near the free surface, pressure transport supplies energy towards the interface, whereas turbulent transport and dissipation are reduced; 
the resulting energy surplus is exported away from the surface predominantly by viscous diffusion. 
The near-surface budget terms do not exhibit a single universal similarity scaling: viscous diffusion is organised over the near-surface viscous scale $\lenvisc$, dissipation over the Kolmogorov sublayer scale $\lenk$, and pressure-related terms require the mixed velocity scale $\ubulk \utau^2 /h$. 
The pressure–strain redistribution further reveals outer–inner coupling: although intense pressure–strain events remain small-scale, their magnitude and directional bias are organised by low-velocity VLSM streaks.
The free-surface influence is therefore best understood as a coupled multi-scale process involving local kinematic constraints, Reynolds-number-dependent surface layers, and outer-layer coherent motions.

\section{Introduction} \label{sec:intro}

Turbulent open channel flow (OCF) is a canonical configuration for many free-surface flows, thus of direct relevance in natural science and engineering applications.
The OCF configuration features a free-slip rigid-lid on the upper boundary, which is much less kinematically constrained than the no-slip wall on the bottom boundary.
Conversely, the closed channel flow configuration (CCF, also known as plane Poiseuille flows) features no-slip walls on both boundaries.
This boundary condition difference significantly modifies turbulence dynamics near the upper boundary, and introduces measurable changes over the full channel depth \citep{bauer2025far}.

The free-slip boundary constrains wall-normal motion only, and it is known to alter the turbulent kinetic energy (TKE) redistribution pathways via the pressure-strain term \citep{Swean1991,Leighton1991,Handler1993,Komori1993,Pan1995}.
If we view this localised modification in the redistribution mechanism as the entry point of the difference between the two channel types, the resulting near-surface effects need to be communicated into and from the channel interior, and it could be possible to trace the energy flow by means of the transport terms in the TKE budget equation.
In this view, the energy budgets can serve as essential diagnostics, since the mean and the second-order statistics alone cannot identify mechanisms.

Mean streamwise velocity distributions in the two channel types are qualitatively similar, except that a clear wake region is absent in OCF \citep{Swean1991}.
Quantitatively, a significantly lower von K\`{a}rm\`{a}n constant in law of the wall was reported for OCF \citep{Pirozzoli2023}.  
Whilst the distributions of wall-normal and the spanwise components of Reynolds normal stresses between the two channel types differ only in the near-surface region, the streamwise component exceeds the corresponding CCF level almost throughout the entire channel depth \citep{Pirozzoli2023,bauer2025far}.

This accumulation of energy in the streamwise component is associated with emergence of very-large-scale motions (VLSMs) \citep{Bauer2015,Duan2020,Duan2021,Pinelli2022}, which were found to be even larger and more energetic than the CCF counterparts \citep{Yao2022,bauer2025far}.
This unique feature demands notably larger numerical domains for OCF to accommodate the enlarged VLSMs.
In CCF, the emergence of VLSMs is responsible for the failure of the wall scaling (i.e. similarity with respect to wall-friction-based velocity) in the streamwise component of Reynolds normal stress \citep{Hoyas2006} and appearance of a mixed-velocity scaling \citep{wei_scaling_2020}.
In OCF, more energetic VLSMs cause the streamwise turbulent intensity to be scaled purely with bulk velocity away from the wall and the free-surface \citep{Bauer2024b}.

By revisiting the pioneering work of \citet{Calmet2003}, our recent contribution \citep{bauer2025far} showed that the surface influence appears through a four-layer structure attached to the free surface.
The thickness of the top two layers is Reynolds-number dependent, implying that a near-surface spatial resolution comparable to the level used near the no-slip wall is required to resolve them truthfully.

Since the inception of the direct numerical simulation of CCF by \citet{Kim1987}, the corresponding TKE budgets have been investigated repeatedly \citep{Mansour1988}.
Despite the long history, however, Reynolds-number dependence of the term-by-term one-point budgets is only sparsely documented \citep[e.g.][]{Hoyas2008}.
Moreover, such investigation is even rarer for the OCF type \citep{Swean1991}, and the budgets near the free surface are particularly under-characterised.
This motivates a systematic budget-level comparison of OCF and CCF across Reynolds number, with particular emphasis on the near-surface region.

Based on the above, we define the following research questions (RQs): 
\begin{enumerate}
    \item Which budget terms transmit the surface influence into the channel interior?
    \item Does a unified similarity scaling exist for the budget terms?
    \item Are pressure-driven inter-component transfer events purely local? 
    Or are they macroscopically organised by outer-layer large-scale motions? 
\end{enumerate}

We address these questions using matched OCF/CCF DNS data \citep{BauerData2023} up to friction Reynolds number $\retau=900$ in a large domain that accommodates VLSMs.
We quantify the full TKE transport budgets, identify term-specific near-surface similarity scalings, and connect pressure–strain redistribution to VLSM organisation via conditional averaging.
Those results can provide useful benchmarks and constraints for future modelling efforts.

The remainder of this paper is organised as follows. 
Section \ref{sec:methods} describes the DNS database, numerical method, and near-surface layer structure used in the analysis. 
Section \ref{sec:results} presents the results, beginning with the OCF/CCF comparison of the TKE budgets (\S \ref{subsec:tke-base}), followed by the similarity analysis of the near-surface budget terms (\S \ref{subsec:similarity}), and the VLSM organisation of pressure–strain redistribution (\S \ref{subsec:pstrain}). 
Section \ref{sec:discussion} discusses the resulting physical picture and its implications for turbulence modelling, and \S \ref{sec:conclusion} concludes the paper.
\section{Methodology}\label{sec:methods}
\subsection{Numerical methods and DNS database} 

This study analyses our open-source DNS database \citep{BauerData2023}, previously investigated in \citet{Bauer2024b,bauer2025far}.
Nevertheless, we summarise the numerical methods and the case specifications for completeness as follows.

The incompressible Navier--Stokes equations are solved in a wall-normal velocity--vorticity formulation using a pseudo-spectral method corresponding to the one employed by \citet{Kim1987}, in which the flow variables are expanded in Fourier series in the homogeneous streamwise ($x$) and spanwise ($z$) directions over equidistant grid points.
In the wall-normal ($y$) direction, the discretisation is based on Chebyshev polynomials on Chebyshev-Gauss-Lobatto points, where the grid resolution is refined towards the no-slip and the free-slip boundaries.
The discretised governing equations are integrated in time with a semi-implicit scheme, where the viscous terms are treated by an implicit Euler approach, whereas a third-order low-storage Runge-Kutta scheme integrates the nonlinear terms explicitly.

By convention, we define a velocity vector $\mathbf{u} = u_i = \begin{bmatrix} u & v & w \end{bmatrix}^\top$ in the above Cartesian coordinate system, as well as pressure $p$.
According to the Reynolds decomposition, the flow variables can be decomposed into the mean component (the temporal and the spatial average in the homogeneous directions) represented by the operator $\langle \phi \rangle$, and the temporal fluctuations with respect to the mean, which is indicated by the superscript $\phi'$, where $\phi$ is an arbitrary variable. 
For example, the above velocity field becomes: $\mathbf{u}(x,y,z,t) = \langle \mathbf{u} \rangle (y) + \mathbf{u}'(x,y,z,t)$.

The computational domain is doubly periodic in $x$- and $z$-directions.
In the OCF set up, a smooth no-slip wall condition is imposed at the lower boundary ($\mathbf{u}=0$ at $y=0$), whilst the free surface is approximated by a free-slip, impermeable surface, enforcing zero wall-normal velocity and wall-normal gradients of the tangential velocity components ($\partial u/\partial y = 0$, $\partial w/\partial y = 0$, $v=0$ at $y=h$, where $h$ is the channel full height in OCF).
In CCF, the upper boundary is substituted by an additional no-slip wall.
A constant volume flow rate is imposed in $x$-direction by adaptive body force.

To characterise the channel flows, two important velocity scales need to be defined, namely the friction velocity $\utau$ and the bulk velocity $\ubulk$ as:

\begin{align}
\utau&=\sqrt{\frac{\langle \tauwall \rangle}{\rho}} \ ,\\
\ubulk&= \frac{1}{h} \int_0^h \langle u \rangle~ \mathrm{d}y \ ,
\end{align}

\noindent
where $\tauwall$ is wall shear stress at the solid boundaries ($\tauwall = \rho \nu(\partial u/\partial y)_\mathrm{wall}$), $\rho$ and $\nu$ are fluid density and kinematic viscosity, $h$ is the channel full height for OCF whilst the semi-height for CCF. 
The database comprises four OCF and four CCF cases at matched friction Reynolds numbers $\retau = \utau h/\nu \approx \{200, 400, 600, 900\}$.
Alternatively, bulk Reynolds numbers $\rebulk = \ubulk h/\nu$ can also be used. 
All cases were simulated in a consistent computational domain of size $L_x = 12\pi h$ and $L_z = 4\pi h$. 
This domain size is sufficient to accommodate VLSMs and to avoid artificial suppression of outer-layer dynamics. 
Details of grid resolution, domain dimensions, Reynolds numbers, and averaging intervals are summarised in table~\ref{tab:cases}.

Statistical quantities are obtained from long-time averaging after the flow has reached a statistically steady-state. 
In addition to one-point statistics accumulated during runtime, full three-dimensional velocity fields are stored at regular intervals and used for the post-processing of second-order moments and TKE budget terms. 
Quantities are normalised either in viscous (wall) units using the friction velocity $\utau$ and viscous length scale $\delta_\nu = \nu/\utau$ being denoted with the superscript $\phi^+$, or in outer units using the bulk velocity $\ubulk$ and $h$, as appropriate.

\begin{table}
\caption{Direct numerical simulation cases analysed in the present study.
All simulations are taken from the DNS database of \citet{BauerData2023}.
}
\label{tab:cases}
\centering
\begin{tabular}{lccccccccccc}
\hline
Case &
$\retau$ &
$\rebulk$ &
$L_x/h$ &
$L_z/h$ &
$N_x$ &
$N_y$ &
$N_z$ &
$\Delta x^+$ &
$\Delta z^+$ &
$\Delta y^+_{\max}$ &
$\Delta T\,\ubulk/h$ \\
\hline
O200 & 200 & 3170  & $12\pi$ & $4\pi$ &  768 & 129 &  512 &  9.8 & 4.9 & 2.5 & 8660 \\
O400 & 399 & 6969  & $12\pi$ & $4\pi$ & 1536 & 193 & 1024 &  9.8 & 4.9 & 3.3 & 1925 \\
O600 & 596 & 11047 & $12\pi$ & $4\pi$ & 1536 & 257 & 1536 & 14.6 & 4.8 & 3.7 & 1460 \\
O900 & 895 & 17512 & $12\pi$ & $4\pi$ & 3072 & 385 & 2048 & 11.0 & 5.5 & 3.7 & 1054 \\
\hline
C200 & 200 & 3170  & $12\pi$ & $4\pi$ &  768 & 129 &  512 &  9.8 & 4.9 & 4.9 & 8600 \\
C400 & 397 & 6969  & $12\pi$ & $4\pi$ & 1536 & 193 & 1024 &  9.8 & 4.9 & 6.5 & 3260 \\
C600 & 593 & 11047 & $12\pi$ & $4\pi$ & 1536 & 257 & 1536 & 14.6 & 4.8 & 7.3 & 1757 \\
C900 & 889 & 17512 & $12\pi$ & $4\pi$ & 3072 & 385 & 2048 & 11.0 & 5.5 & 7.3 & 1013 \\
\hline
\end{tabular}
\end{table}

\subsection{Near-surface multi-layer structure}

As it will be shown, the OCF--CCF discrepancies are confined to the vicinity of the free surface. 
To organise this region and to facilitate Reynolds-number comparisons, we briefly summarise the near-surface multi-layer structure which we refer throughout the remainder of this study. 

In our recent contribution \citep{bauer2025far} we showed that the surface influence appears through a four-layer structure attached to the free surface, consisting of: \textit{surface-influenced layer} $\deltasil$, \textit{normal-velocity-dumping layer} $\deltanvd$, \textit{near-surface viscous sublayer} $\deltavisc$, and \textit{Kolmogorov sublayer} $\deltak$, listed in descending order of thickness.
First, the surface-influenced layer characterises the extent of the surface influence in the streamwise component of TKE, and it is in order of the full channel depth ($\deltasil \approx h$).
Second, the normal-velocity-dumping layer, characterising the deviation of $v_\rms \equiv \sqrt{\langle v'v'\rangle}$ from the CCF profile, is much thinner than the first layer ($\deltanvd \approx 0.3h$, e.g. \citet{Nagaosa1999,Duan2021}).
Third, the near-surface viscous sublayer is where $v_\rms$ grows linearly as moving away from the free surface, whose thickness is characterised by a viscous characteristic length scale $\lenvisc \equiv h \rebulk^{-1/2}$, as $\deltavisc \approx \lenvisc$.
Finally, the Kolmogorov sublayer, which is the inner-most layer and characterised by another length scale $\lenk \equiv h \rebulk^{-3/4}$ with $\deltak \approx 20\lenk$, is where the surface-normal gradients of the mean velocity ($d\langle u \rangle/dy$) as well as the intensity of the surface-parallel vorticity components ($\omega_{x/z,\rms}$) are damped to zero.
\section{Results} \label{sec:results}
\subsection{TKE budget and baseline comparison}
\label{subsec:tke-base}

To address RQ1 defined in \S \ref{sec:intro}, namely: ``\textit{Which budget terms transmit the surface influence into the channel interior?}'', we analyse the TKE budget in the viscous normalisation for each Reynolds number. 
The comparison is made between open- and closed-channel configurations at matched $\retau$. 
Generally, some degree of OCF--CCF discrepancies are expected due to the difference in the boundary condition: in OCF the surface-normal component of turbulent kinetic energy vanishes at $y=h$ to satisfy the impermeable boundary condition, which also facilitates inter-component energy transfer from the surface-normal to the surface-parallel directions via the pressure-strain term in the component-wise budget equations \citep{Swean1991,Leighton1991,Handler1993,Komori1993,Pan1995}.
Conversely, impermeability is not enforced at the channel semi-height in CCF, therefore such energy transfer is not relevant.
Note that this pressure–strain redistribution term in the component budgets is distinct from the pressure transport in the TKE budget presented below. 
Due to its importance, however, the pressure-strain term will be further analysed in a separate section.

The turbulent kinetic energy budget for fully-developed channel flows can be written as:
\begin{equation}
\underbrace{\nu \frac{\mathrm{d}^2 k}{\mathrm{d} y^2}}_{\text{(i)}} 
\underbrace{-\frac{1}{\rho}\frac{\mathrm{d} \langle v'p' \rangle}{\mathrm{d} y}}_{\text{(ii)}}
\underbrace{-\frac{1}{2}\frac{\mathrm{d}\langle v'u'_ju'_j \rangle}{\mathrm{d}y}}_{\text{(iii)}}
\underbrace{-\langle u'v' \rangle \frac{\mathrm{d} \langle u \rangle}{\mathrm{d} y}}_{\mathcal{P}}
\underbrace{-\nu \left\langle \left(\frac{\partial u'_i}{\partial x_j}\right)^2\right\rangle}_{-\tilde{\varepsilon}}
=0\ , 
\label{eq:tkebud}
\end{equation}
where $k=\langle u'_i u'_i \rangle /2$ defines TKE, (i-iii) represent the transport terms, namely: (i) viscous diffusion, (ii) pressure transport, and (iii) turbulent transport, whereas $\mathcal{P}$ and $\tilde{\varepsilon}$ are production and (pseudo-)dissipation, respectively.

Notice that the above TKE budget equation is in the pseudo-dissipation form.
In the present statistically one-dimensional configuration, true dissipation $\varepsilon$ differs from $\tilde{\varepsilon}$ by a wall-normal viscous-transport correction involving $\mathrm{d}^2\langle v'^2\rangle/\mathrm{d}y^2$. 
Rewriting the budget in terms of true dissipation therefore preserves the total balance but redistributes part of the near-surface viscous contribution between dissipation and diffusion. 
Since this regrouping combines terms associated with different near-surface length scales, the pseudo-dissipation form is retained for the following term-wise similarity analysis.
For simplicity, we refer to $\tilde{\varepsilon}$ as \textit{dissipation} in the rest of this paper.

\begin{figure} 
\centering
\includegraphics[]{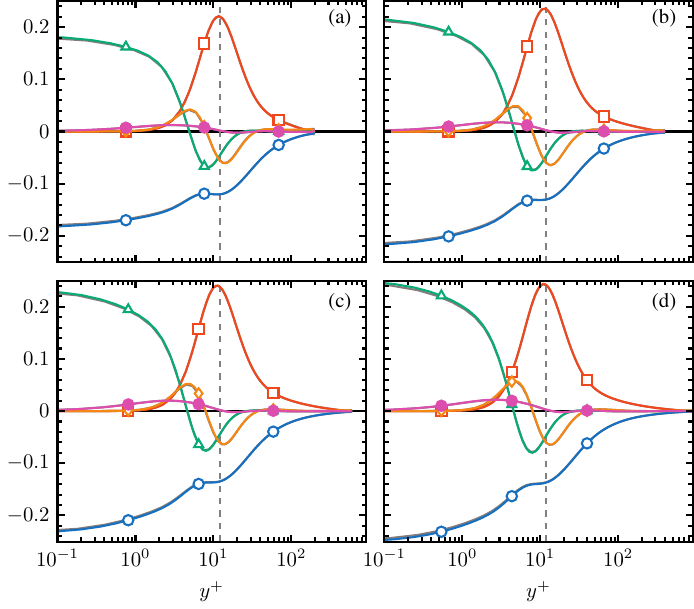}
\caption{TKE budget for OCF in the near-wall region. 
\symbb, production; \symba, pseudo-dissipation; \symbc, viscous diffusion; \symbd, turbulent transport; \symbe, pressure transport. 
The magnitudes of the terms are normalised by $\utau^4/\nu$, whereas the normalised distance from the no-slip wall is $y^+$. 
(a) O200; (b) O400; (c) O600; (d) O900. 
Grey lines indicate the corresponding CCF data.
}
\label{fig:tkewall}
\end{figure}

\begin{figure} 
\centering
\includegraphics[]{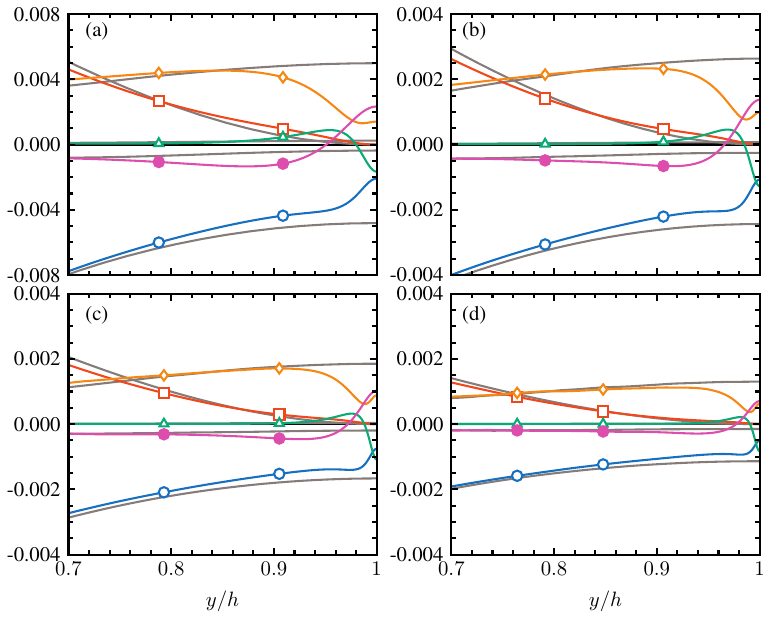}
\caption{TKE budget for OCF in the near-surface region as a function of free surface distance. \symbb, production; \symba, pseudo-dissipation; \symbc, viscous diffusion; \symbd, turbulent transport; \symbe, pressure transport. 
The magnitudes of the terms are normalised by $\utau^4/\nu$, whereas the vertical positions are normalised by the outer scale $h$. 
(a) O200; (b) O400; (c) O600; (d) O900. 
Grey lines indicate the corresponding CCF data. 
Notice a difference in the vertical axis limits in (a) compared to the rest.}
\label{fig:tkesurf}
\end{figure}

Figure \ref{fig:tkewall} depicts the TKE budget terms normalised by $\utau^4/\nu$ as functions of $y^+$. 
The wall-normal coordinate is shown on a logarithmic scale to emphasise the near-wall region and the channel interior.
Figure \ref{fig:tkesurf} shows the same budget terms in a near-surface representation, using a linear $y/h$ coordinate and a smaller spatial range.

In figure \ref{fig:tkewall}, it is visible that the OCF budget term profiles generally match their CCF counterparts very closely in the vicinity of the wall and throughout most of the flow domain, to the extent that the CCF lines are barely visible in the graph. 
Conversely, in the near-surface region approximately above $y/h \approx 0.7$ shown in figure \ref{fig:tkesurf}, the profiles between the two channel types differ fundamentally.
It appears that all the budget terms except the production term are qualitatively altered.
%

The positive peak in the pressure transport profile directly at the free surface shown in figure \ref{fig:tkesurf} is a clear signature of the altered pressure-induced transport, since it is entirely absent in the CCF counterpart, which indicates that a substantial amount of energy is transported from the interior towards the free surface.

The turbulent transport term in this region also deviates from the CCF level significantly as the magnitude drops notably.
Similarly, the magnitude of the dissipation term drops sharply as the near-surface velocity field adapts to the free-slip boundary condition.
This local reduction of dissipation results in an excess of energy to be transported away from the free surface, which is done predominantly by viscous diffusion.
Conversely, the same viscous diffusion term is entirely inactive outside of the near-surface region.

As mentioned, the production term remains qualitatively similar between the two channel types even in the limit $y \rightarrow h$.
By definition, the production term needs to disappear at $y=h$ in both configurations, since a kinematic condition $\mathrm{d}\langle u\rangle/\mathrm{d}y=0$ is enforced by the free-slip boundary condition (OCF), and by the geometric symmetry (CCF).
Despite the similarity, however, the production appears to be consistently higher in OCF slightly below the free surface before dropping to zero, whereas the opposite is true in the channel interior. 
The corresponding crossover points between the two channel types appear to be highly Reynolds number dependent.
The observed trend can be attributed to a localised increase of the mean shear slightly below the free surface before converging to zero, which was first discussed by \cite{Shen2000}.
Subsequently, our recent study showed that the local mean shear peak appears at the outer edge of so-called Kolmogorov sublayer \citep[cf.][Fig. 2]{bauer2025far}.

\begin{figure}
    \centering
    \includegraphics[]{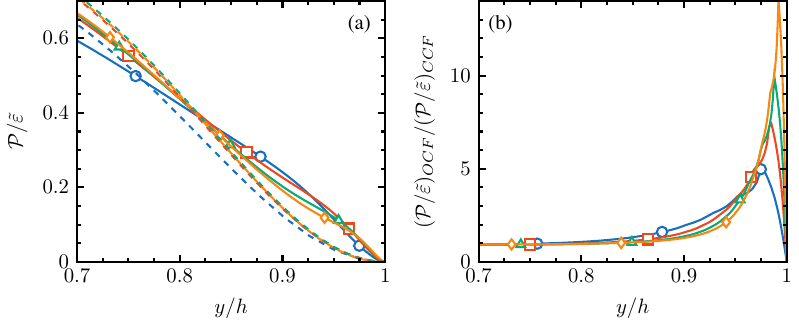}
    \caption{Near-surface production–dissipation ratio $\mathcal{P}/\tilde{\varepsilon}$ for OCF (solid lines) and CCF (dashed lines) (a) the ratio $[\mathcal{P}/\tilde{\varepsilon}]_{\mathrm{OCF}}/[\mathcal{P}/\tilde{\varepsilon}]_{\mathrm{CCF}}$ (b). \symbb, $\mathrm{Re}_\tau\approx200$; \symba, $\mathrm{Re}_\tau\approx400$; \symbc, $\mathrm{Re}_\tau\approx600$; \symbd, $\mathrm{Re}_\tau\approx900$.}
    \label{fig:prod_diss_ratio}
\end{figure}

To assess whether the observed OCF/CCF differences in the production term translate into a materially different local balance, we additionally examined the ratio $\mathcal{P}/\tilde{\varepsilon}$ in the near-surface region (cf. figure \ref{fig:prod_diss_ratio}).
Before focusing on the near-surface region, we confirmed that the usual approximate local-equilibrium behaviour is recovered in the logarithmic layer: at $\retau > 200$, $\mathcal{P}/\tilde{\varepsilon}$ approaches unity in both OCF and CCF (not shown).
The behaviour shown in figure \ref{fig:prod_diss_ratio} therefore represents the breakdown of this
interior production--dissipation balance as the free surface is approached.
In both OCF and CCF, $\mathcal{P}/\tilde{\varepsilon}$ remains below unity throughout $0.7\leq y/h \leq 1$ and decreases towards zero in the limit $y \rightarrow h$, consistent with the requirement that production vanishes at the free-slip boundary (OCF) and at the symmetry plane (CCF). 
Moreover, the ratio-of-ratios $[\mathcal{P}/\tilde{\varepsilon}]_{\mathrm{OCF}}/[\mathcal{P}/\tilde{\varepsilon}]_{\mathrm{CCF}}$ stays close to unity over most of the interval and only rises considerably in the immediate vicinity of the surface where both numerator and denominator become small. 
This indicates that, whilst $\mathcal{P}$ exhibits a modest OCF/CCF deviation slightly below the free surface, the local production-dissipation partitioning does not provide an additional discriminating mechanism beyond what is already evident from the transport terms.
%

In summary, the free-surface influence enters the TKE budget through the kinematic constraint at $y=h$, which suppresses wall-normal motions and induces a strong near-surface inter-component energy redistribution.
Meanwhile, the vertical energy flux towards the free surface due to pressure transport increases significantly, whilst the local dissipation and turbulent transport decrease.
Consequently, this results in a near-surface energy surplus that must be exported into the interior. 
This energy export away from the surface is realised predominantly by viscous diffusion, which becomes the dominant transport mechanism in the immediate vicinity of the free surface whilst remaining negligible outside the surface layer. 
\subsection{Surface-layer similarity of transport terms}\label{subsec:similarity}

To address RQ2: ``\textit{Does a unified similarity scaling exist for the budget terms?}'', here we investigate whether the near-surface budget term profiles exhibit a Reynolds-number similarity when expressed in interface-based coordinates. 
In particular, we seek a scaling for the surface-normal distance and the transport-term amplitudes that makes the OCF profiles across the Reynolds numbers similar.

\begin{figure} 
\includegraphics[width=0.5\textwidth]{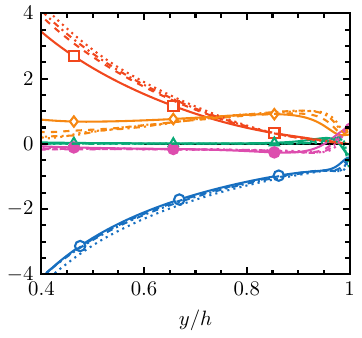}
\centering
\caption{TKE budget for open channel flow as a function of distance from the wall normalised with the channel height $h$. 
The amplitudes of the budget terms are normalised by $\utau^3/h$. 
Solid (\blackline) lines, O200; dashed (\blackdashed) lines, O400; dash-dotted (\blackdashdotted) lines, O600; dotted (\blackdotted) lines, O900. 
\symbb, production; \symba, pseudo-dissipation; 
\symbc, viscous diffusion; 
\symbd, turbulent transport; 
\symbe, pressure transport. 
}
\label{fig:tkescaling}
\end{figure}

Before focusing on the near-surface similarity, however, we need to establish another baseline by evaluating whether the outer flow similarity observed for CCF applies to the current OCF data.
For CCF, \citet{Hoyas2008} showed that all the TKE budget terms can be made similar under the $(\utau^3/h)$-scaling.
Since $\retau = \utau h/\nu$, this normalisation by $\utau^3/h$ is equivalent to multiplying the wall-scaled budget amplitude by $\retau$.
Correspondingly, figure \ref{fig:tkescaling} presents the TKE budget terms in OCF normalised by this scaling in their amplitude, whilst the vertical locations are scaled by $h$.
Apart from the expected low-Reynolds-number effect, which are mainly visible on the production and turbulent transport terms, all terms scale well in the outer flow region away from the wall and the free surface. 

\begin{figure} 
\includegraphics[width=\textwidth]{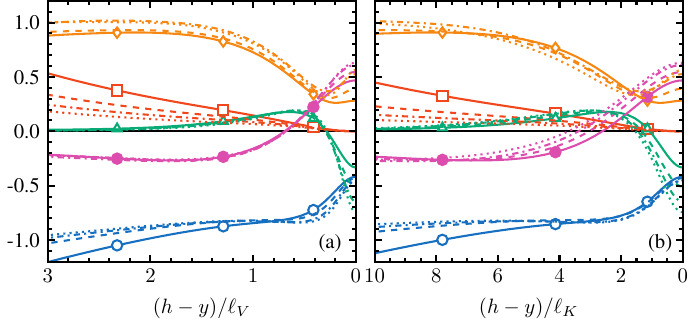}
\centering
\caption{TKE budget for open channel flow as a function of free surface distance. 
The amplitudes of the budget terms are normalised by $\utau^3/h$. 
Solid (\blackline) lines, O200; dashed (\blackdashed) lines, O400; dash-dotted (\blackdashdotted) lines, O600; dotted (\blackdotted) lines, O900. 
\symbb, production; \symba, pseudo-dissipation; 
\symbc, viscous diffusion; 
\symbd, turbulent transport; 
\symbe, pressure transport. 
The distance from the free surface is normalised with (a) the near-surface viscous length scale $\lenvisc$ and (b) the near-surface Kolmogorov length scale $\lenk$.
}
\label{fig:tkescalingb}
\end{figure}

Focusing on the near-surface region, it appears that the viscous diffusion and the pressure transport profiles become similar in the region slightly away from the free surface when the surface distances are normalised by the near-surface viscous scale $\lenvisc$ (cf. figure \ref{fig:tkescalingb}a, $(h-y)/\lenvisc \geq 0.4$).
In contrast, Reynolds number dependencies in the amplitudes of those terms remain in the immediate vicinity of the free surface, signalling that different scales control the amplitudes in this region.

The dissipation profiles in the direct vicinity of the free surface, on the other hand, can be made similar when the free-surface distances are normalised by the Kolmogorov scale $\lenk$ whilst keeping the amplitude scaling as before (cf. figure \ref{fig:tkescalingb}b, $(h-y)/\lenk \leq 4$).
The observed trend is consistent with a conjecture made by \citet{Calmet2003} that the near-surface dissipation should be linked to the local enstrophy, whilst our previous work \citep{bauer2025far} showed that the variation length scale of the quantity is indeed $\lenk$.

\begin{figure} 
\includegraphics[width=\textwidth]{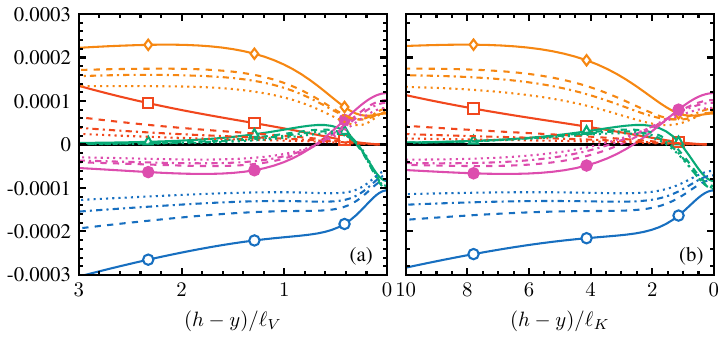}
\centering
\caption{
TKE budget for open channel flow as a function of free surface distance. Here normalised by $\ubulk^3/h$. 
Solid (\blackline) lines, O200; dashed (\blackdashed) lines, O400; dash-dotted (\blackdashdotted) lines, O600; dotted (\blackdotted) lines, O900. 
\symbb, production; 
\symba, pseudo-dissipation; 
\symbc, viscous diffusion; 
\symbd, turbulent transport; 
\symbe, pressure transport.
The distance from the free surface is normalised with (a) the near-surface viscous length scale $\lenvisc$; 
(b) the near-surface Kolmogorov length scale $\lenk$.}
\label{fig:tkescaling2}
\end{figure}

In figure \ref{fig:tkescaling2}, the amplitudes of the budget terms are normalised by $\ubulk^3/h$.
With this bulk scaling, a clear similarity is obtained for the viscous diffusion term in the immediate proximity of the free surface $(h-y)/\lenvisc \leq 0.4$.
This amplitude scaling property can be attributed to the emergence of significantly more energetic VLSMs in OCF compared with CCF.
In \cite{Bauer2024b}, we showed that OCF VLSMs appear at Reynolds numbers as low as $\retau=400$, causing $u_\rms \equiv \sqrt{\langle u'u' \rangle}$ to scale with $\ubulk$ for $y \gtrsim 0.3h$. 
Consistently, the component-wise budget analysis in Appendix~\ref{app:uu-budget} shows that the viscous diffusion term in the $\langle u'u'\rangle$-budget collapses well under $\ubulk^3/h$ scaling. 
The length scale controlling the near-surface profile variation remains $\lenvisc$.

The above combined amplitude and length scaling can also be interpreted from the balance between a viscous
diffusion time scale and an advective time scale determining the thickness of the interfacial boundary layer. 
The viscous diffusion time over a wall-normal distance $\ell$ is $\tau_\nu \sim \ell^2/\nu$. 
Equating this with the advective time scale $\tau_\mathrm{adv}\sim h/\ubulk$, which is assumed to scale with the bulk quantities due to the presence of VLSMs, gives
\begin{equation}
\ell \sim \left(\frac{\nu h}{\ubulk}\right)^{1/2}
= h \rebulk^{-1/2}
= \lenvisc \ .
\end{equation}
If the surface-parallel turbulent energy participating in this balance is set by the
VLSM-induced streamwise fluctuations, it should be of order $\ubulk^2$ and the following estimate can be made
\begin{equation}
\nu \frac{\mathrm{d}^2 k}{\mathrm{d} y^2} \sim \nu \frac{\ubulk^2}{\ell_V^2}
\sim \frac{\ubulk^3}{h}.
\end{equation}
Hence, this order-of-magnitude estimate explains why the viscous diffusion term collapses with $\lenvisc$ in distance and $\ubulk^3/h$ in amplitude.

\begin{figure} 
\includegraphics[width=\textwidth]{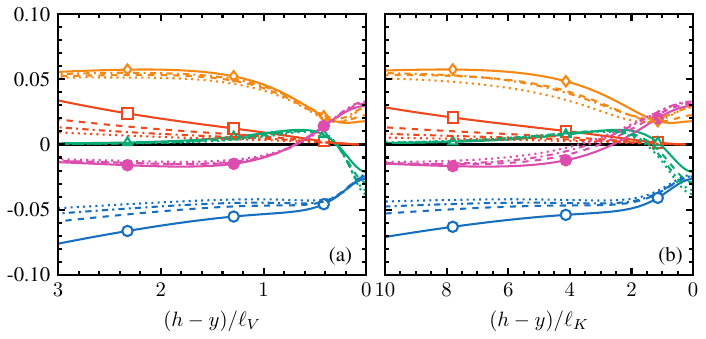}
\centering
\caption{
TKE budget for open channel flow as a function of free surface distance. Here normalised by $\ubulk \utau^2/h$. 
Solid (\blackline) lines, O200; dashed (\blackdashed) lines, O400; dash-dotted (\blackdashdotted) lines, O600; dotted (\blackdotted) lines, O900. 
\symbb, production; 
\symba, pseudo-dissipation; 
\symbc, viscous diffusion; 
\symbd, turbulent transport; 
\symbe, pressure transport.
The distance from the free surface is normalised with (a) the near-surface viscous length scale $\lenvisc$; 
(b) the near-surface Kolmogorov length scale $\lenk$.}
\label{fig:tkescaling4}
\end{figure}

As visible in figures \ref{fig:tkescaling}, \ref{fig:tkescalingb}, and \ref{fig:tkescaling2}, the amplitudes of the pressure transport term cannot be made similar either by $\utau^3/h$, or $\ubulk^3/h$. 
Instead, a significantly improved collapse is obtained when the term is normalised by a mixed velocity scale, namely $\ubulk \utau^2/h$, as shown in figure \ref{fig:tkescaling4}.
A possible origin of this scaling trend will be discussed in the subsequent subsection.

Whilst we could successfully identify the appropriate scales for the rest of the transport terms and the dissipation term, the turbulent transport term exhibits only partial similarity under the same set of normalisations.
Among the tested scales, the mixed $(\ubulk \utau^2/h)$-scaling provides the most competitive reduction of Reynolds-number dependence when it is combined with $\lenvisc$-normalised free surface distance, however a non-negligible residual separation remains.
In particular, in the immediate vicinity of the free surface, $(h-y)/\lenvisc \leq 0.4$, the residual Reynolds number dependence is the most notable.
The corresponding component-wise Reynolds-normal-stress budgets, reported in Appendix \ref{app:uu-budget}, show that this partial similarity of the TKE turbulent-transport term reflects heterogeneous scaling behaviour among the individual normal-stress components.

In summary, the budget terms in the free-surface region \textit{do not} exhibit a \textit{single} near-surface similarity scaling.
Instead, the similarity is both distance-specific $\lenvisc$ versus $\lenk$ and velocity-scale-specific ($\utau$, $\ubulk$, or mixed). 
In particular, dissipation collapses on $\lenk$, whereas viscous diffusion and pressure transport are controlled by $\lenvisc$; however, their amplitudes scale differently, with pressure transport requiring the mixed velocity scale $\ubulk\utau^2/h$.
To organise these key results, the identified scaling properties of the individual terms are summarised in table \ref{tab:TKE-budget-summary}.
%
\begin{table}
    \centering
    {
    \begin{tabular}{c c c c c c}
         &  viscous (i) & pressure (ii) & turbulent (iii) & production ($\mathcal{P}$) & dissipation ($-\tilde{\varepsilon}$) \\
         \hline
         near-surface & $\ubulk^3/h$, $\lenvisc$ & $\ubulk \utau^2 / h$, $\lenvisc$ & partial & -- & $\utau^3/h$, $\lenk$ \\
         \hline
         bulk (interior) & $\utau^3/h$, $h$ & $\utau^3/h$, $h$ & $\utau^3/h$, $h$ & $\utau^3/h$, $h$ & $\utau^3/h$, $h$ \\
         \hline
    \end{tabular}
    }
    \caption{Result table of identified scales of TKE budget terms}
    \label{tab:TKE-budget-summary}
\end{table}
\subsection{VLSM organisation of near-surface pressure-strain}
\label{subsec:pstrain}
Here, we address RQ3: ``\textit{Are pressure-driven inter-component transfer events purely local? Or are they macroscopically organised by outer-layer large-scale motions?}'', focusing primarily on the OCF configuration where the VLSMs are more energetic and directly relevant to the near-surface dynamics.

The pressure-strain tensor is defined as:
\begin{equation}
 \Pi_{ij}=\left\langle \frac{p'}{\rho}\left(\frac{\partial u'_i}{\partial x_j}+\frac{\partial u'_j}{\partial x_i}\right)\right\rangle\ , 
\end{equation}
    
\noindent
whose trace sums up to zero due to incompressibility, and it does not appear in equation \ref{eq:tkebud}, as mentioned.
Still, the pressure-strain term plays an important role in the overall dynamics of turbulent kinetic energy, as the diagonal terms redistribute the energy between the components.
In OCF, this mechanism transfers the surface-normal component of turbulent kinetic energy ($\langle v'v'\rangle$) into the surface-parallel components ($\langle u'u'\rangle$, $\langle w'w'\rangle$) below the free surface, resulting an increased anisotropy in the region (\cite{Calmet2003} termed it as \textit{mechanism II}).
Conversely, the same mechanism was found to be responsible for reducing the anisotropy of the normal stresses generally above the buffer layer in CCF \citep{Mansour1988,Hoyas2008}, which is commonly referred to as \textit{return-to-isotropy} effect (\textit{mechanism I} in \cite{Calmet2003}).
This return-to-isotropy effect is also significant in OCF away from the wall and the free surface, as demonstrated below.

\begin{figure} 
\centering
\includegraphics[width=\textwidth]{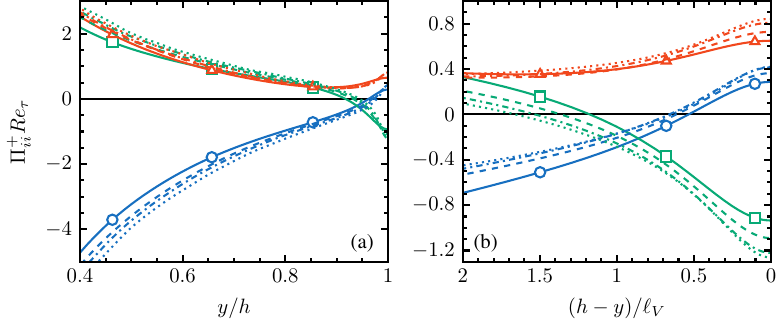}
\caption{Diagonal components of the pressure-strain term, here normalised by $\utau^3/h$.
Solid (\blackline) lines, O200; dashed (\blackdashed) lines, O400; dash-dotted (\blackdashdotted) lines, O600; dotted (\blackdotted) lines, O900.
\symbx, $\Pi_{11}$; \symby, $\Pi_{22}$; \symbz, $\Pi_{33}$. 
The distance from the free surface is normalised with (a) the channel height $h$; (b) the near-surface viscous length scale $\lenvisc$.}
\label{fig:pstrainscaling}
\end{figure}

In figure \ref{fig:pstrainscaling}, the profiles of the diagonal components of the pressure-strain tensor are shown.
The amplitudes of the diagonal terms are normalised by the basic $(\utau^3/h)$-scaling, which leads to a satisfactory collapse throughout the visible vertical extent.

In the channel interior ($y/h < 0.85$), it is evident from negative $\Pi_{11}$ and positive $\Pi_{22/33}$ that the return-to-isotropy effect is active as expected, redistributing energy from the streamwise direction to the spanwise and the surface-normal directions.
A subtle Reynolds number dependency in the amplitudes can be observed, which we attribute to the emergence of VLSMs in the current data at around $\retau=400$ \citep{Bauer2024b}, progressively biasing the energy content into the streamwise direction with increasing Reynolds number, making the turbulent stresses more anisotropic as the result.
Recall that the emergence VLSMs and the increased anisotropy is not a unique feature of OCF, as it also appears in CCF  \citep{Hoyas2006}, though it only becomes energetically significant from Reynolds numbers much higher than the above limit.
The pressure-strain term responds to the increased anisotropy, and based on the magnitudes of $\Pi_{22/33}$ in figure \ref{fig:pstrainscaling}, it appears that the energy is redistributed to the surface-normal and the spanwise directions fairly equally.

In contrast, the surface-normal energy is redistributed to the surface-parallel components within $y/h \geq 0.85$, which is evident from negative $\Pi_{22}$ and the corresponding positive $\Pi_{11/33}$.
Based on the magnitudes, it is clear that the energy is redistributed more in the spanwise component than in the streamwise counterpart, independent of Reynolds number.
Whilst this directional bias is consistent with the findings in the earlier low-Reynolds number DNS studies \citep{Swean1991,Komori1993,Nagaosa1999}, it contradicts with a conjecture by \citet{Calmet2003}, who stated that the ratio between the spanwise and streamwise pressure-strain terms should approach to unity at high Reynolds numbers.
A plausible contributor for this contradiction is that the above statement was made based on the higher degree of isotropy observed in their LES data, which was simulated in a computational domain that is too small for VLSMs to exist \citep{Bauer2024b}.
Beside the uncertainties linked to the use of a sub-grid-scale model, the contribution of the above limitation is supported by the data that the magnitude of their streamwise Reynolds stress component \citep{Calmet2003} lies notably below the corresponding OCF data of this study, as well as the CCF data of \citet{Hoyas2006}.
As the results of a more recent LES study in highly restricted numerical domains \citep{Ahmed02012021} coincide with this trend, it indicates a persistent nature of this particular constraint. 

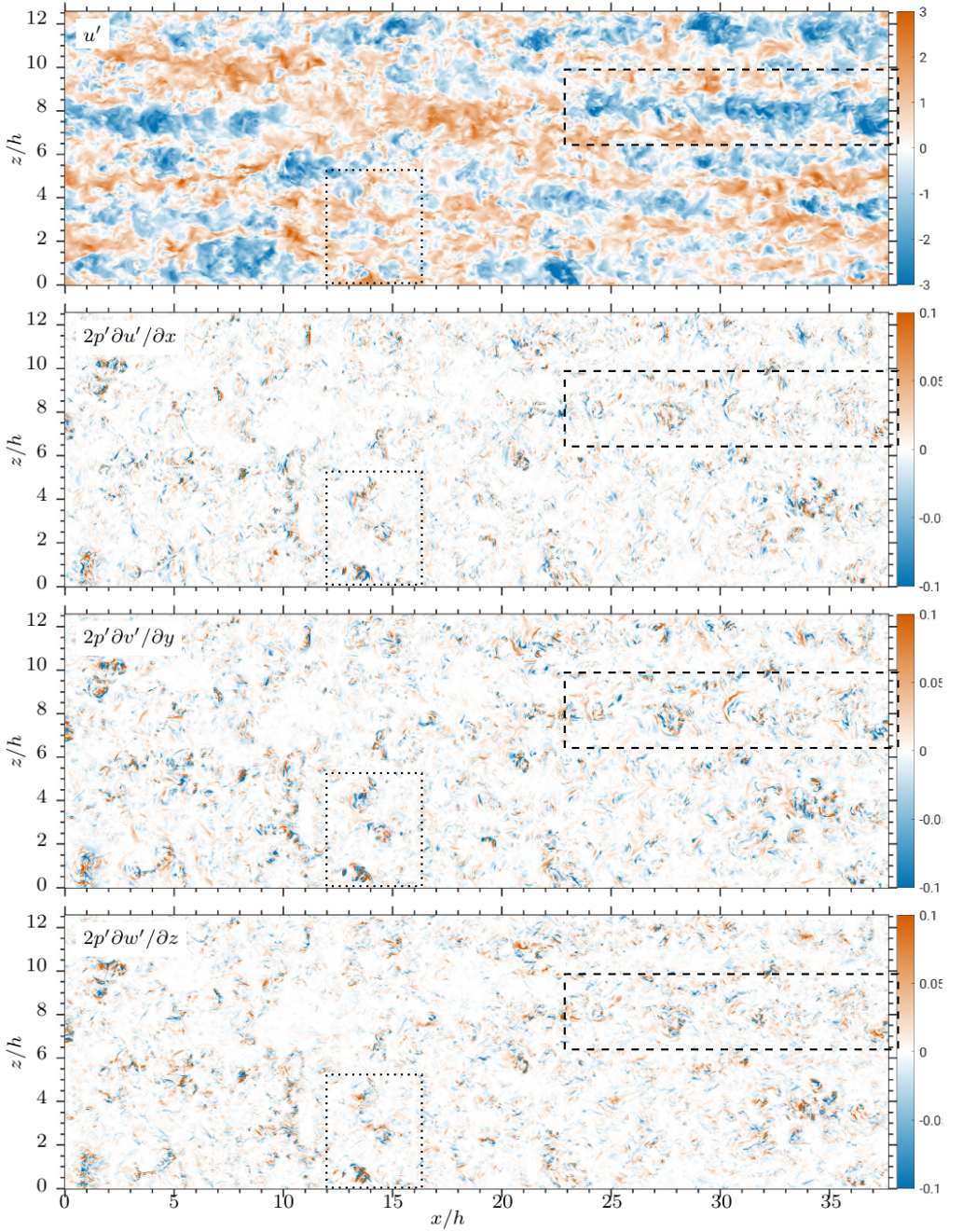
\begin{figure}
    \centering
    \begin{tikzpicture}
        \node[inner sep=0, anchor=south west] (img) at (0,0) 
        {\includegraphics[width=\linewidth]{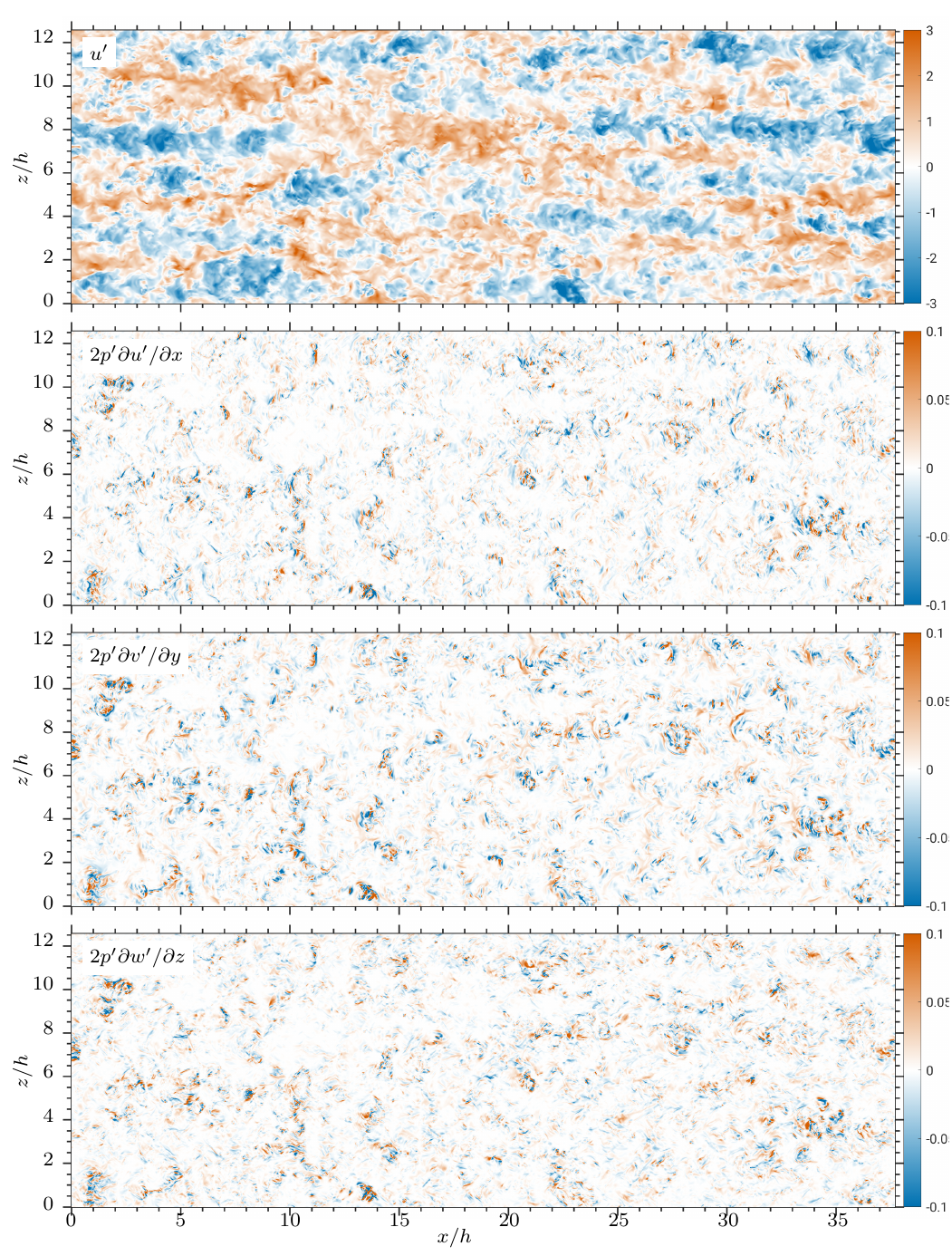}};

        \begin{scope}[x={(img.south east)}, y={(img.north west)}] 
        \draw[black, dashed, line width=0.8pt] (0.60,0.87) rectangle (.95,.93); 
        \draw[black, dashed, line width=0.8pt] (0.60,0.63) rectangle (.95,.69); 
        \draw[black, dashed, line width=0.8pt] (0.60,0.39) rectangle (.95,.45); 
        \draw[black, dashed, line width=0.8pt] (0.60,0.15) rectangle (.95,.21); 

        \draw[black, dotted, line width=0.8pt] (0.35,0.76) rectangle (.45,.85); 
        \draw[black, dotted, line width=0.8pt] (0.35,0.52) rectangle (.45,.61); 
        \draw[black, dotted, line width=0.8pt] (0.35,0.28) rectangle (.45,.37); 
        \draw[black, dotted, line width=0.8pt] (0.35,0.04) rectangle (.45,.13); 
        \end{scope}
    \end{tikzpicture}
    \caption{Instantaneous: (a) streamwise velocity fluctuation $u'$, and (b-d) pressure-strain components from the same instance. 
    OC600 at $y/h=0.95$.
    }
    \label{fig:vlsms_pressure_strain}
\end{figure}

Next, we examine potential significance of VLSMs on the pressure-strain terms, directly addressing RQ3.
Figure \ref{fig:vlsms_pressure_strain} depicts a near-surface instantaneous streamwise velocity fluctuation ($u'$) field primarily visualising VLSMs, and the corresponding normal components of the instantaneous pressure-strain products.
Recall that the channel-flow VLSMs generally emerge in a form of large-scale low-velocity streaks (cf. the highly-elongated blue regions in figure \ref{fig:vlsms_pressure_strain}), associated with underlying large-scale streamwise vortices, commonly referred to as super-streamwise vortices (SSVs) \citep{Zhong2016,Duan2021,scherer:21a,bauer2025far}.

Distributions of the individual components of the pressure-strain products clearly show that they are explicitly coupled (cf. the region marked by a dotted-line rectangle, ranging from $x/h \approx 12$) as expected.
Moreover, whilst the high-intensity pressure-strain events themselves are small-scale phenomena, which is evident in the finely detailed individual structures, it appears that their positions are macroscopically organised by the large-scale low-velocity streaks (cf. the region marked by a dashed-line rectangle, ranging from $x/h \approx 23$).

\begin{figure} 
\centering
\includegraphics[]{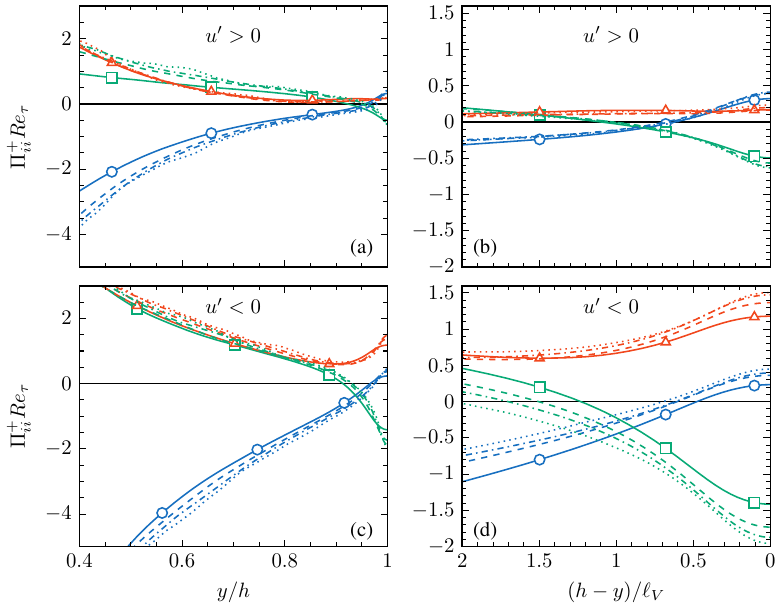}
	\caption{Conditional average of the diagonal components of the pressure-strain term $\Pi_{ii}$, here normalised by $\utau^3/h$. Legend see figure~\ref{fig:pstrainscaling}. (a,b) $u'>0$; (c,d) $u'<0$. The distance from the free surface is normalised with (a,c) the channel height $h$; (b,d) the near-surface viscous length scale $\lenvisc$.}
\label{fig:pstraincond}
\end{figure}

To confirm, figure \ref{fig:pstraincond} shows the profiles of the pressure-strain products conditionally averaged with respect to the sign of the underlying $u'$.
As expected from the visual impression in figure \ref{fig:vlsms_pressure_strain}, the magnitudes of the pressure-strain terms are found to be significantly larger ($\approx 3\times$ at the surface) within the low-velocity streak regions.
Moreover, notice that the profile in the spanwise component inside the high-velocity streaks does not feature a distinct localised increase towards the free surface, but it remains approximately constant throughout the near-surface region instead (cf. figure \ref{fig:pstraincond}b).
This indicates that the surface-normal energy component in the high-velocity streaks is preferentially redistributed towards the streamwise direction (cf. the blue lines above the orange lines in figure \ref{fig:pstraincond}b), whereas in the low-velocity streak region the energy is transferred mainly to the spanwise component (figure \ref{fig:pstraincond}d), being consistent with the statistics without conditional averaging (figure \ref{fig:vlsms_pressure_strain}).

\begin{figure}
    \centering
    \includegraphics[width=0.8\linewidth]{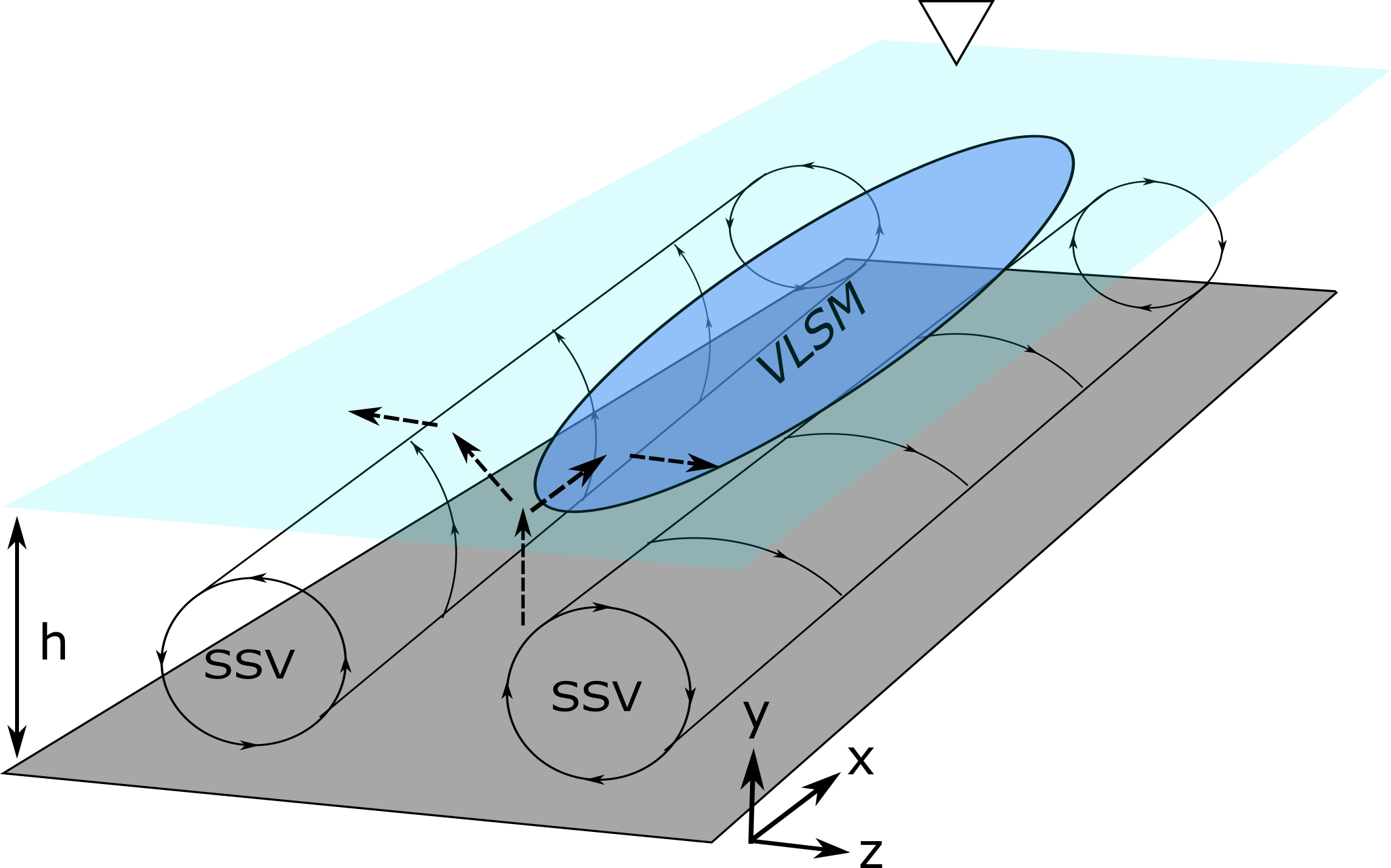}
    \caption{Conceptual sketch of super-streamwise vortex preferentially inducing inter-component energy transfer in the spanwise direction.
    Arrows with dashed line represent the preferential directions of energy redistribution at different heights.
    }
    \label{fig:ssv-vlsm}
\end{figure}

This observed qualitative difference in the dominant direction of the energy redistribution can be linked to the underlying near-surface vortical structures and their interaction with the free surface.
As mentioned, VLSMs are known to coexist with the large-scale super-streamwise vortices (SSVs) spanning over the entire channel height, and we suppose that those SSVs residing inside the VLSMs guide the fluctuating velocity in the direction perpendicular to the rotational axis on the free surface plane (i.e. $z$-direction, cf. figure \ref{fig:ssv-vlsm} for a conceptual sketch).
On the other hand, inside the large-scale high-velocity streaks, the current data indicate that the dominant vortical motions are around the spanwise axis, guiding the surface velocity fluctuations in the streamwise direction.
The above conceptual model was inspired by, and analogous to the streamwise vortical tubes and the associated splats/anti-splats discussed in \citet[][Fig. 10, 11]{Nagaosa1999}.
Here, it is important to point out a key difference: \citet{Nagaosa1999} simulated OCF at a marginal Reynolds number of $\retau=150$, in which scales are not separated (i.e. no VLSMs) and their near-wall buffer-layer turbulent structures spanned over $h$, interacting directly with the free surface instead. 

\begin{figure} 
\centering
\includegraphics[width=\textwidth]{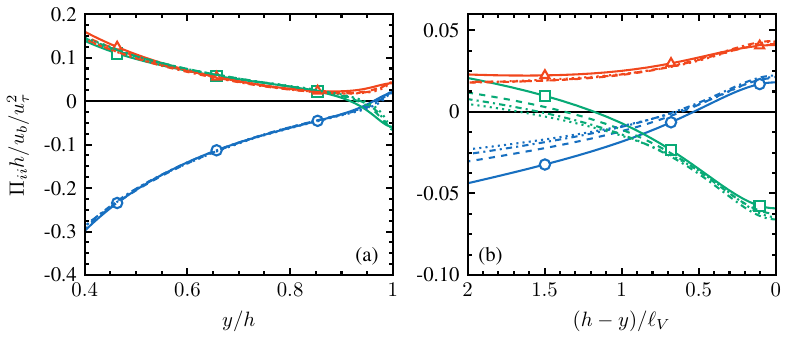}
\caption{Diagonal components of the pressure-strain term, here normalised by $\ubulk\utau^2/h$. 
Solid (\blackline) lines, O200; dashed (\blackdashed) lines, O400; dash-dotted (\blackdashdotted) lines, O600; dotted (\blackdotted) lines, O900.
\symbx, $\Pi_{11}$; \symby, $\Pi_{22}$; \symbz, $\Pi_{33}$. 
The distance from the free surface is normalised with (a) the channel height $h$; (b) the near-surface viscous length scale $\lenvisc$.}
\label{fig:pstrainscaling4}
\end{figure}

Finally, we focus on the similarity in the pressure-strain components.
Figure \ref{fig:pstrainscaling4} depicts the same pressure-strain term profiles under the mixed-velocity scaling of $\ubulk\utau^2/h$, which was determined for the pressure transport term (cf. \S \ref{subsec:similarity}).
Under this normalisation, the similarity of the profiles both in the interior as well as the near-surface region improves drastically. 
The most appropriate variation length scale of the profiles seems to be $\lenvisc$, as for the pressure transport term.

\begin{figure} 
\centering
\includegraphics[width=.6\textwidth]{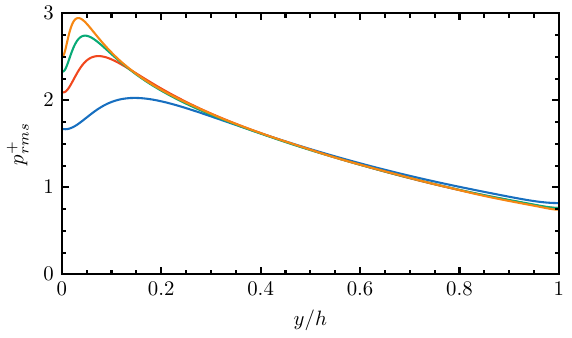}
\caption{OCF $p_\mathrm{rms}$ normalised by $\utau^2$. \linea, O200; \lineb, O400; \linec, O600; \lined, O900.
}
\label{fig:prms}
\end{figure}

This scaling behaviour can be explained by examining the definition of the streamwise component of the pressure-strain term: $2\langle p'\partial u'/\partial x \rangle$ (note: $\rho$ is omitted since it is constant).
Whilst the root-mean-square of pressure fluctuation scales with $\utau^2$ away from the wall (cf. figure \ref{fig:prms}), the amplitude of $u'$ scales with $\ubulk$ and the variation length scale is $h$ at the free surface due to the emergence of VLSMs, leading to the $(\ubulk\utau^2/h)$-scaling.
Since all the diagonal components are tightly coupled, the $2\langle p'\partial v'/\partial y \rangle$ and the $2 \langle p'\partial w'/\partial z \rangle$ components should also scale in $\ubulk\utau^2/h$.
Additional pressure--strain scalings are provided in Appendix \ref{app:p-strain-additional} to document the comparison with alternative velocity normalisations.

In summary, the near-surface pressure–strain field is not only a local redistribution mechanism imposed by the free-surface boundary condition, but is also organised by outer-layer coherent motions. 
In the channel interior, pressure–strain retains its usual return-to-isotropy role, redistributing energy from the streamwise component towards the spanwise and surface-normal components. 
Near the free surface, this direction reverses: surface-normal energy is transferred into the surface-parallel components, with a persistent bias towards the spanwise direction. 
Instantaneous fields and conditional averages show that this redistribution is strongly amplified within low-velocity VLSM streaks, where the magnitude of the pressure–strain terms is substantially larger than in high-speed regions. 
The conditional statistics further suggest a streak-dependent redistribution pathway: low-velocity streaks favour transfer into the spanwise component, whereas high-speed regions show relatively stronger transfer into the streamwise component. 
This behaviour is consistent with a picture in which VLSM-associated large-scale vortical motions organise the otherwise small-scale pressure–strain events. 
Finally, the pressure–strain profiles collapse substantially better under the mixed velocity scaling $\ubulk\utau^2/h$ and the surface-normal length scale $\lenvisc$, consistent with the corresponding pressure-transport scaling.


\section{Discussion}\label{sec:discussion}

The results above suggest that the free-surface influence should not be viewed as a purely local boundary adjustment. 
Instead, the near-surface budget behaviour reflects an interaction between the kinematic constraints at the interface, Reynolds-number-dependent surface-layer scales, and the energetic outer-layer motions characteristic of OCF.

A perhaps counter-intuitive outcome is that viscous diffusion becomes the dominant transport mechanism in the limit $y \rightarrow h$. 
The dominance of viscous diffusion near the free surface should therefore be interpreted together with the Reynolds-number-dependent surface-layer structure. 
The budget analysis in \S \ref{subsec:tke-base} identifies viscous diffusion as the principal outward transport pathway for the surface-induced TKE surplus, whilst the scaling analysis in \S \ref{subsec:similarity} shows that this pathway is organised over the near-surface viscous length scale $\lenvisc$. 
Thus, the near-surface viscous layer is not only visible in the wall-normal velocity statistics reported previously, but also appears directly in the TKE budget as the finite-thickness region through which molecular diffusion exports energy away from the interface. 
In contrast, the dissipation adjustment occurs on the smaller Kolmogorov sublayer scale $\lenk$, indicating that transport and dissipation respond to the free-surface constraint on distinct Reynolds-number-dependent length scales.

The conditional pressure–strain statistics further indicate that the redistribution process cannot be regarded as a purely local near-surface adjustment. 
Although the instantaneous pressure–strain events remain small-scale, their occurrence and directional bias are organised by the large-scale streak structure. 
This provides a direct link between the enhanced streamwise energy content associated with OCF VLSMs and the near-surface inter-component energy transfer imposed by the free-surface constraint.
In this sense, the pressure–strain term acts as a bridge between outer-layer organisation and local surface-layer anisotropy.

The Reynolds-number dependence observed in the near-surface TKE budget also has implications for turbulence modelling. 
%
The free-surface region cannot be treated as a trivial counterpart of either the near-wall region or the symmetry plane of a closed channel. 
The present results show that near-surface transport, dissipation, and pressure redistribution are governed by distinct length and velocity scales: viscous diffusion is organised over the near-surface viscous scale $\lenvisc$, dissipation adjusts over the smaller Kolmogorov sublayer scale $\lenk$, and pressure-related terms involve a mixed bulk/friction-velocity scaling. 
This poses a serious challenge for closures based on a single local turbulence length scale or a single velocity scale. 
In particular, eddy-viscosity RANS closures and conventional LES subgrid-scale models may reproduce the mean flow or second-order statistics whilst still misrepresenting the partitioning of TKE transport near the surface, especially when pressure transport and pressure–strain redistribution are not modelled explicitly. 
The present DNS results therefore provide useful constraints for free-surface turbulence closures: models should account for the Reynolds-number-dependent surface-layer thickness, the non-universal scaling of the individual budget terms, and the organisation of pressure-driven redistribution by large-scale coherent motions.
\section{Conclusion} \label{sec:conclusion}

We analysed the turbulent kinetic energy (TKE) budget and pressure--strain redistribution in matched open- and closed-channel DNS up to $\retau \approx 900$, using a computational domain sufficiently large to accommodate the energetic VLSMs characteristic of OCF. 
The aim was to determine how the free-surface influence is transported into the channel interior, whether the near-surface budget terms admit a unified similarity scaling, and whether pressure-driven inter-component transfer is organised by outer-layer coherent motions.

The free-surface influence is found to be communicated primarily through the transport terms of the TKE budget and through pressure--strain redistribution in the component-wise energy balance. 
Near the free surface, pressure transport supplies energy towards the interface, while turbulent transport and dissipation are reduced. 
The resulting energy surplus is exported away from the surface predominantly by viscous diffusion, which becomes the dominant outward transport pathway within the immediate surface layer. 
The corresponding similarity analysis shows that this region does not admit a single universal scaling: viscous diffusion is organised over the near-surface viscous scale $\lenvisc$, dissipation adjusts over the smaller Kolmogorov sublayer scale $\lenk$, and pressure-related terms require the mixed velocity scale $\ubulk \utau^2/h$.

The pressure--strain results further show that inter-component redistribution near the free surface is not purely local. 
Away from the surface, pressure--strain retains its usual return-to-isotropy role, whereas near the surface it transfers energy from the surface-normal component into the surface-parallel components, with a persistent spanwise bias. 
Instantaneous fields and conditional averages demonstrate that this redistribution is amplified in low-velocity VLSM streaks and that its directional character depends on the underlying large-scale streak structure.

Overall, the free-surface influence is best understood as a coupled multi-scale process: local kinematic constraints suppress wall-normal motion and turbulent transport, Reynolds-number-dependent surface layers set the transport and dissipation length scales, and VLSMs organise the pressure-related redistribution. 
These findings provide benchmark constraints for free-surface turbulence modelling, particularly for closures based on a single local turbulence scale or closures that do not explicitly represent pressure transport and pressure--strain redistribution.
\vspace{1.5cm}

\noindent{\bf  Supplementary data\bf{.}} \label{SupMat} The data supporting the findings of this study are available at: \url{https: //doi . o rg/1 0.41 21/88 678f0 2-2a34 -445 2-8534-6361fc34d06b}. \\

\noindent{\bf Acknowledgements\bf{.}} 
The authors thank Dr. M. Uhlmann for his important role in
initiating the DNS database used in this study, for supporting the earlier open-channel-flow
work from which the present study developed, and for helpful scientific discussions.
The authors also gratefully acknowledge the initial funding support provided through the
DFG project led by Dr. M. Uhlmann, under which parts of the database and earlier
related work were developed.
Y.S. also thanks Dr. M. Manhart for a fruitful discussion and valuable suggestions in the interpretation of scaling behaviours.
During preparation of this manuscript, the corresponding author used DeepL and ChatGPT (OpenAI) solely to identify potential English-language issues (e.g. grammar, phrasing, and readability) and to suggest alternative wording. 
These tools were not used to generate scientific content, analyse data, create figures, or draw conclusions. 
All changes were reviewed and approved by the corresponding author, who takes full responsibility for the manuscript content.\\

\noindent{\bf Funding\bf{.}} The authors gratefully acknowledge the financial support of the DFG under grant no. UH~242/3-1. 
The simulations were carried out at the HPC clusters UC2 and CARA. Thus, 
the authors gratefully acknowledge the scientific support and HPC resources provided by the SCC Karlsruhe, the German Aerospace Center (DLR), and the state of Baden-Wuerttemberg through bwHPC.
The HPC system CARA is partially funded by ``Saxon State Ministry for Economic Affairs, Labour and Transport'' and ``Federal Ministry for Economic Affairs and Climate Action''.
\\

\noindent{\bf Declaration of Interests}. The authors report no conflict of interest. \\

\noindent{\bf  Author ORCID\bf{.}}  
Y. Sakai, \url{https://orcid.org/0000-0001-8125-6566}; \\ 
C. Bauer, \url{https://orcid.org/0000-0003-1838-6194}\\

\noindent{\bf Author contributions\bf{.}} 
Both authors contributed to the study conception and design.
C.B. performed the numerical simulations and post-processing. 
Y.S. wrote the first draft of the manuscript, and both authors contributed to revisions. 
Both authors approved the final manuscript.

\begin{appendix}
\clearpage
\section{Additional scaling of TKE budget terms} \label{app:tke-additional}

\begin{figure} 
\includegraphics[width=\textwidth]{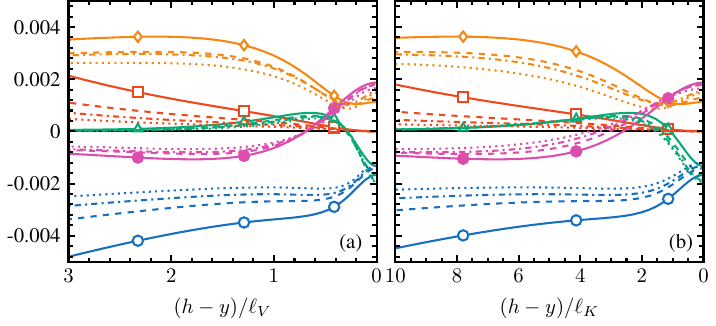}
\centering
\caption{
TKE budget for open channel flow as a function of free surface distance. Here normalised by $\ubulk^2 \utau/h$. 
Solid (\blackline) lines, O200; dashed (\blackdashed) lines, O400; dash-dotted (\blackdashdotted) lines, O600; dotted (\blackdotted) lines, O900. 
\symbb, production; 
\symba, pseudo-dissipation; 
\symbc, viscous diffusion; 
\symbd, turbulent diffusion; 
\symbe, pressure diffusion.
The distance from the free surface is normalised with (a) the near-surface viscous length scale $\lenvisc$; 
(b) the near-surface Kolmogorov length scale $\lenk$.}
\label{fig:tkescaling3}
\end{figure}

Here, we present an additional mixed velocity scaling for the TKE budget terms in figure \ref{fig:tkescaling3}, for completeness.

\clearpage
\section{Scaling of $\langle u_i'u_i' \rangle$ budget terms} \label{app:uu-budget}
Here, we present the distributions of the terms in the $\langle u_i'u_i' \rangle$ budget equations under different normalisations.
The steady-state Reynolds-stress budget equation in the present configuration reads:

\begin{equation}
    0 = P_{ij} + \Pi_{ij} - \tilde{\varepsilon}_{ij} + D^{\nu}_{ij} + D^{\mathrm{t}}_{ij} + D^{p}_{ij}
    \label{eq:reynodls-stress-bud}
\end{equation}

\noindent
where $P_{ij}$ represents production, $\Pi_{ij}$ is pressure-strain, and $\tilde{\varepsilon}_{ij}$ denotes pseudo dissipation.
Furthermore, $D^{\nu}_{ij}$, $D^{\mathrm{t}}_{ij}$ and $D^{p}_{ij}$ are the viscous diffusion, turbulent transport, and pressure transport terms, respectively.
Individual definitions of these terms are as follows:

\begin{equation}
\begin{split}
P_{ij} &= -\left(\left(\langle u_i'v'\rangle\delta_{j1} + \langle u_j'v'\rangle\delta_{i1}\right) \frac{\mathrm{d} \langle u \rangle}{\mathrm{d} y}\right),\\
\Pi_{ij} &= \left\langle \frac{p'}{\rho}\left( \frac{\partial u'_i}{\partial x_j} + \frac{\partial u'_j}{\partial x_i} \right) \right\rangle , \\
\tilde{\varepsilon}_{ij} &= 2\nu \left\langle \frac{\partial u'_i}{\partial x_k}\frac{\partial u'_j}{\partial x_k} \right\rangle , \\
D^{\nu}_{ij} &= \nu \frac{\mathrm{d}^2 \langle u'_i u'_j \rangle }{\mathrm{d} y^2} , \\
D^{\mathrm{t}}_{ij} &= -\frac{\mathrm{d}\langle v'u'_i u'_j \rangle}{\mathrm{d}y} , \\
D^{p}_{ij} &= -\frac{1}{\rho}\left( \frac{\mathrm{d} \langle u_i'p' \rangle}{\mathrm{d} y}\delta_{j2} + \frac{\mathrm{d} \langle u_j'p' \rangle}{\mathrm{d} y}\delta_{i2}\right)\ .
\end{split}
\end{equation}

Focusing on the normal stress components (i.e. $i=j$), the production term is only active in the streamwise component $\langle u'u'\rangle$, whereas the pressure transport term is only active in the surface-(wall-)normal component $\langle v'v'\rangle$ by definition.
As discussed in the main text, the pressure-strain term is responsible for inter-component energy transfer.

\begin{figure} 
\centering
\includegraphics[]{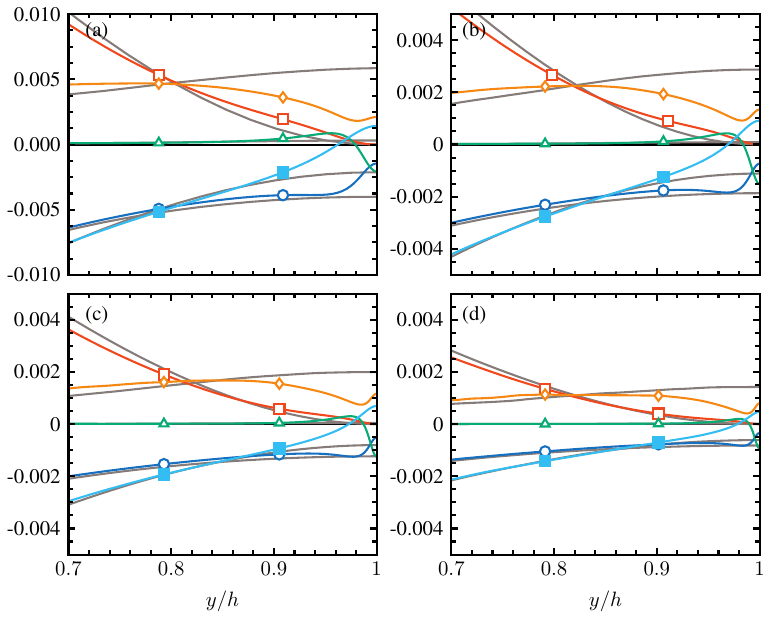}
\caption{$\langle u'u'\rangle$ budget for OCF in the near-surface region as a function of free surface distance. \symbb, production; \symba, pseudo-dissipation; \symbc, viscous diffusion; \symbd, turbulent transport; \symbf, pressure-strain.
The magnitudes of the terms are normalised by $\utau^4/\nu$, whereas the vertical positions are normalised by the outer scale $h$. 
(a) O200; (b) O400; (c) O600; (d) O900. 
Grey lines indicate the corresponding CCF data. 
Notice a difference in the vertical axis limits in (a) compared to the rest.}
\label{fig:uusurf}
\end{figure}
\begin{figure} 
\centering
\includegraphics[]{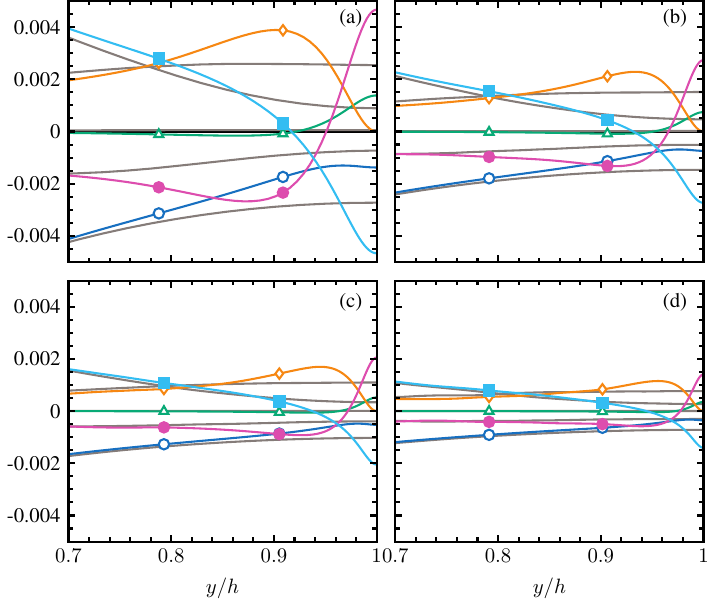}
\caption{$\langle v'v'\rangle$ budget for OCF in the near-surface region as a function of free surface distance. \symba, pseudo-dissipation; \symbc, viscous diffusion; \symbd, turbulent transport; \symbe, pressure transport; \symbf, pressure-strain.
The magnitudes of the terms are normalised by $\utau^4/\nu$, whereas the vertical positions are normalised by the outer scale $h$. 
(a) O200; (b) O400; (c) O600; (d) O900. 
Grey lines indicate the corresponding CCF data. 
Notice a difference in the vertical axis limits in (a) compared to the rest.}
\label{fig:vvsurf}
\end{figure}
\begin{figure} 
\centering
\includegraphics[]{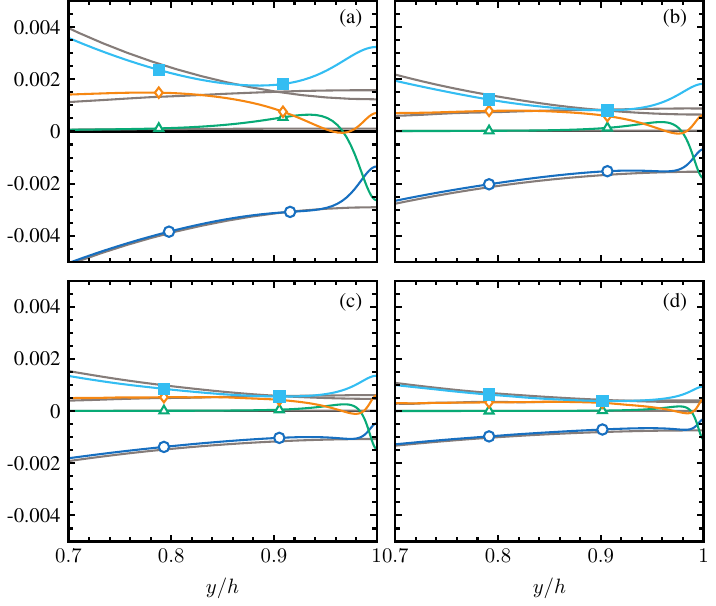}
\caption{$\langle w'w'\rangle$ budget for OCF in the near-surface region as a function of free surface distance. \symba, pseudo-dissipation; \symbc, viscous diffusion; \symbd, turbulent transport; \symbf, pressure-strain.
The magnitudes of the terms are normalised by $\utau^4/\nu$, whereas the vertical positions are normalised by the outer scale $h$. 
(a) O200; (b) O400; (c) O600; (d) O900. 
Grey lines indicate the corresponding CCF data. 
Notice a difference in the vertical axis limits in (a) compared to the rest.}
\label{fig:wwsurf}
\end{figure}

Figures \ref{fig:uusurf}, \ref{fig:vvsurf} and \ref{fig:wwsurf} depict the budget terms of the Reynolds normal stresses normalised by $\utau^4/\nu$ as a function of $y/h$.
The coloured lines are from OCF, whereas the grey lines are from CCF.

The distributions of the $\langle u'u'\rangle$-budget terms, shown in figure \ref{fig:uusurf}, largely resemble the ones from the TKE budget: the production profile, which is slightly elevated from the CCF baseline, combined with a sharp reduction in the dissipation, creates localised energy component surplus in the near-surface region, which needs to be transported away from the free surface predominantly by the viscous diffusion term.
Though the level is significantly lower than the CCF counterpart, the turbulent diffusion term also contributes positively to the near-surface energy component balance by transporting the normal stress from the channel interior.
Additionally, the positive inter-component energy transfer from the surface-normal component by the pressure-strain term contributes to the underlying local energy component surplus, which needs to be countered solely by the viscous term.

A similar observation can be made for the $\langle w'w'\rangle$-budget (cf. figure \ref{fig:wwsurf}), except the fact that the production is absent in this component, which implies that the energy component needs to be transferred from somewhere else to exist.
As discussed in the main text, the pressure-strain term is responsible for this, converting noticeably more surface-normal energy to the spanwise component than to the streamwise counterpart.
Note that the reported sharp reductions of $\varepsilon_{11}$ and $\varepsilon_{33}$ in the close proximity of the free surface are consistent with the observations of \citet{Handler1993}.

The $\langle v'v'\rangle$-budget depicted by figure \ref{fig:vvsurf} shows a qualitatively different picture from the surface-parallel counterparts. 
The production term is again inactive in this component, whilst the pressure-strain term is a negative (sink) contributor to the energy component balance in this direction. 
This implies that there is no \textit{in-situ} energy component source in the near-surface region, thus it needs to be transported from the channel interior by other means.
The largest transporter in this case is the pressure (diffusion) transport term, whilst the turbulent transport and the viscous diffusion terms also contribute positively to the local energy component balance.
On the other hand, negative (sink) contribution by the dissipation term is not negligible but moderate, whilst most of the energy component is removed from the near-surface region by the pressure-strain term, and distributed to the surface-parallel components, as discussed.
Furthermore, the viscous diffusion term $D^\nu_{22}$ and the dissipation rate $\tilde{\varepsilon}_{22}$ in balance at the free surface, whilst the pressure transport and pressure-strain terms ($D^p_{22}$ and $\Pi_{22}$) cancel each other also at the interface.
Since these observations hold independently of $\retau$, the present results support and extend the corresponding findings of \cite{Handler1993}, which were obtained at $\retau=125$.

In figure \ref{fig:uusurf}, it appears that the free surface influence in the production and the turbulent transport terms of the $\langle u'u'\rangle$-budget persists deep inside the channel interior even at higher Reynolds numbers, whereas the OCF-CCF discrepancies in the rest of the terms diminish with increasing $\retau$ in the channel interior.
Conversely, the OCF-CCF discrepancies generally disappear at higher Reynolds numbers for $\langle v'v' \rangle$- (cf. figure \ref{fig:vvsurf}) and $\langle w'w'\rangle$- (cf. figure \ref{fig:wwsurf}) budget terms.
Since the footprint of VLSMs mainly appear in $\langle u'u'\rangle$, the persistent OCF-CCF discrepancies in the corresponding production and the turbulent transport budget terms can be attributed to the emergence of the highly energetic OCF VLSMs.

\begin{figure} 
\includegraphics[width=0.9\textwidth]{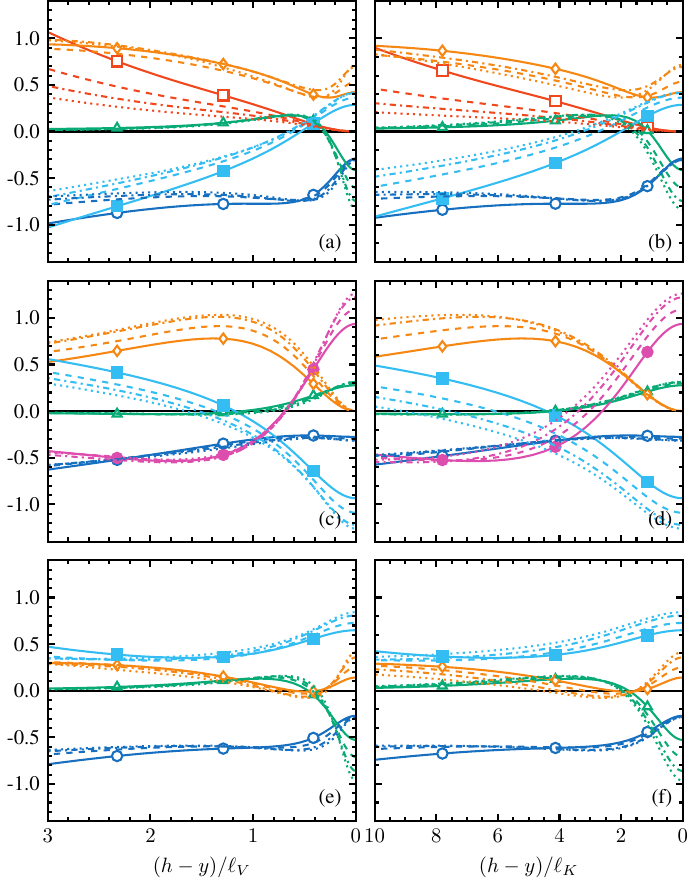}
\centering
\caption{$\langle u_i'u_i'\rangle$ budget for open channel flow as a function of free surface distance. (a,b) $\langle u'u'\rangle$; (c,d) $\langle v'v'\rangle$; (e,f) $\langle w'w'\rangle$.
Here normalised by $\utau^3/h$. 
Solid (\blackline) lines, O200; dashed (\blackdashed) lines, O400; dash-dotted (\blackdashdotted) lines, O600; dotted (\blackdotted) lines, O900. 
\symbb, production; \symba, pseudo-dissipation; 
\symbc, viscous diffusion; 
\symbd, turbulent transport; 
\symbe, pressure transport;
\symbf, pressure-strain. 
The distance from the free surface is normalised with 
 (a,c,e) the near-surface viscous length scale $\lenvisc$; 
 (b,d,f) the near-surface Kolmogorov length scale $\lenk$.}
\label{fig:uuscaling}
\end{figure}

\begin{figure} 
\includegraphics[width=0.9\textwidth]{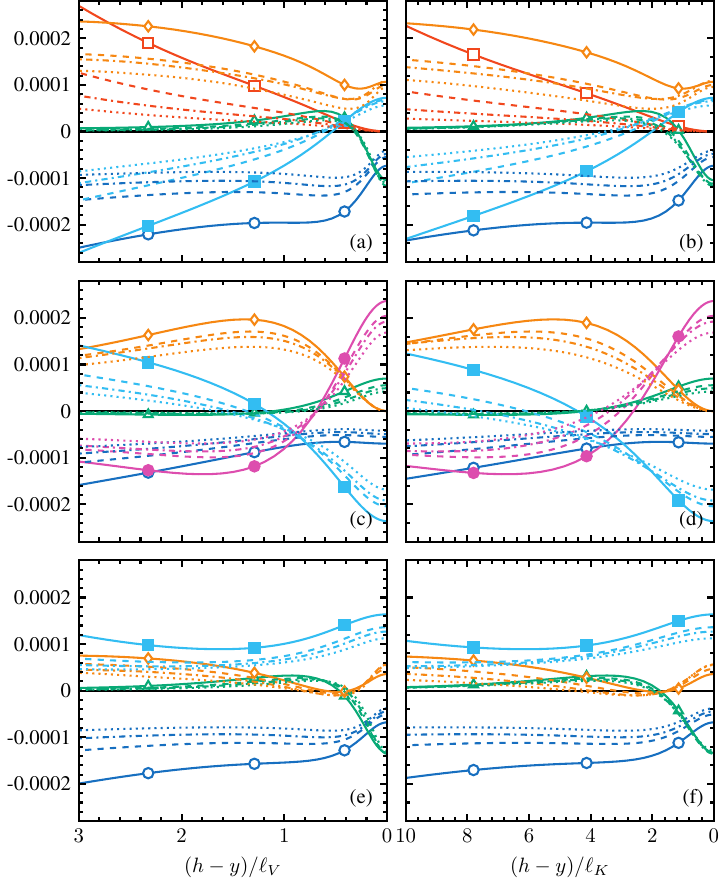}
\centering
\caption{$\langle u_i'u_i'\rangle$ budget for open channel flow as a function of free surface distance. (a,b) $\langle u'u'\rangle$; (c,d) $\langle v'v'\rangle$; (e,f) $\langle w'w'\rangle$.
Here normalised by $\ubulk^3/h$. 
Solid (\blackline) lines, O200; dashed (\blackdashed) lines, O400; dash-dotted (\blackdashdotted) lines, O600; dotted (\blackdotted) lines, O900. 
\symbb, production; \symba, pseudo-dissipation; 
\symbc, viscous diffusion; 
\symbd, turbulent transport; 
\symbe, pressure transport;
\symbf, pressure-strain. 
The distance from the free surface is normalised with 
 (a,c,e) the near-surface viscous length scale $\lenvisc$; 
 (b,d,f) the near-surface Kolmogorov length scale $\lenk$.}
\label{fig:uuscaling2}
\end{figure}

\begin{figure} 
\includegraphics[width=0.9\textwidth]{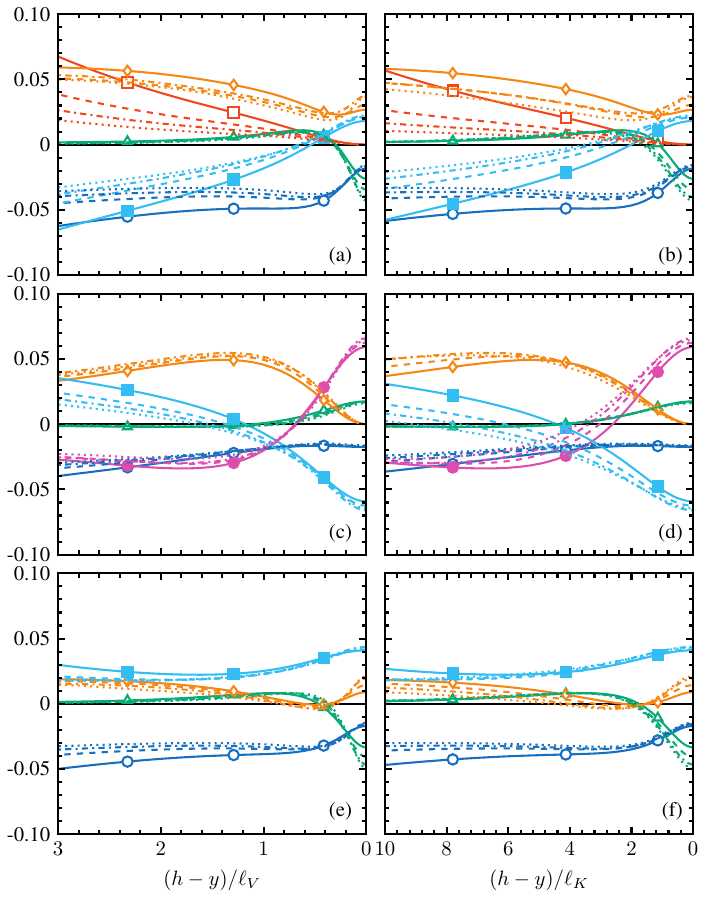}
\centering
\caption{$\langle u_i'u_i'\rangle$ budget for open channel flow as a function of free surface distance. (a,b) $\langle u'u'\rangle$; (c,d) $\langle v'v'\rangle$; (e,f) $\langle w'w'\rangle$.
Here normalised by $\ubulk\utau^2/h$. 
Solid (\blackline) lines, O200; dashed (\blackdashed) lines, O400; dash-dotted (\blackdashdotted) lines, O600; dotted (\blackdotted) lines, O900. 
\symbb, production; \symba, pseudo-dissipation; 
\symbc, viscous diffusion; 
\symbd, turbulent transport; 
\symbe, pressure transport. 
\symbf, pressure-strain. 
The distance from the free surface is normalised with 
 (a,c,e) the near-surface viscous length scale $\lenvisc$; 
 (b,d,f) the near-surface Kolmogorov length scale $\lenk$.}
\label{fig:uuscaling4}
\end{figure}

\begin{figure} 
\includegraphics[width=0.9\textwidth]{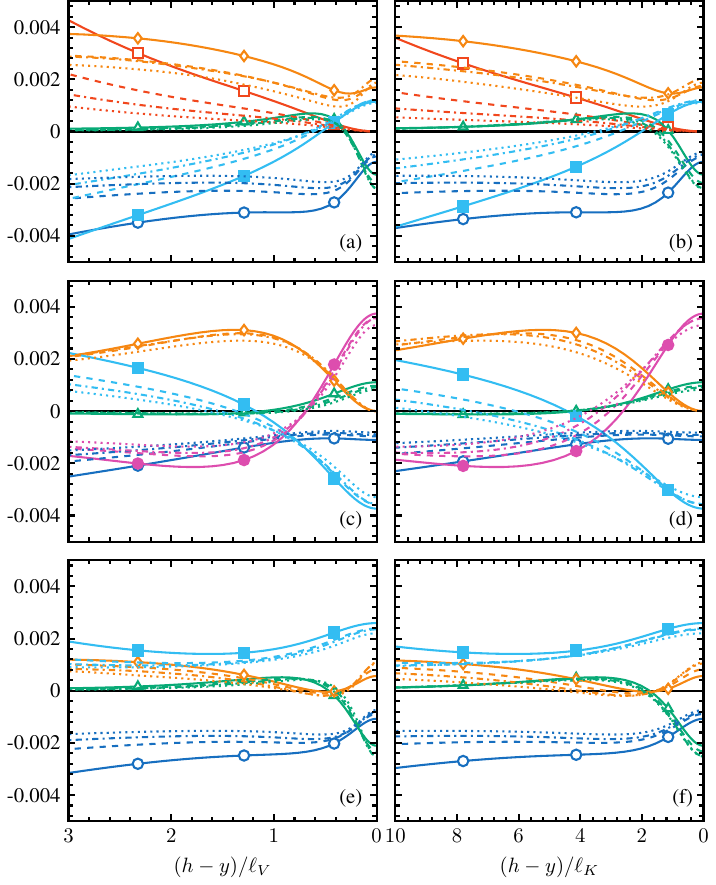}
\centering
\caption{$\langle u_i'u_i'\rangle$ budget for open channel flow as a function of free surface distance. (a,b) $\langle u'u'\rangle$; (c,d) $\langle v'v'\rangle$; (e,f) $\langle w'w'\rangle$.
Here normalised by $\ubulk^2\utau/h$. 
Solid (\blackline) lines, O200; dashed (\blackdashed) lines, O400; dash-dotted (\blackdashdotted) lines, O600; dotted (\blackdotted) lines, O900. 
\symbb, production; \symba, pseudo-dissipation; 
\symbc, viscous diffusion; 
\symbd, turbulent transport; 
\symbe, pressure transport.
\symbf, pressure-strain. 
The distance from the free surface is normalised with 
 (a,c,e) the near-surface viscous length scale $\lenvisc$; 
 (b,d,f) the near-surface Kolmogorov length scale $\lenk$.}
\label{fig:uuscaling3}
\end{figure}

Figure \ref{fig:uuscaling} shows the $\langle u'_iu'_i\rangle$ budget terms from the OCF cases, where the near-surface profiles from all four Reynolds numbers are plotted together. 
Here, the amplitude of each budget term is normalised with $\utau^3/h$, whereas the free surface distance is normalised by $\lenvisc$ and $\lenk$.
The same profiles but in different amplitude normalisations are shown in: figures \ref{fig:uuscaling2} with $\ubulk^3/h$, \ref{fig:uuscaling4} with $\ubulk\utau^2/h$, and \ref{fig:uuscaling3} with $\ubulk^2\utau/h$ normalisations, respectively.

As for the TKE budget, the dissipation terms in the $\langle u_i'u_i'\rangle$ budgets in the free surface region collapse very well when they are normalised by $\utau^3/h$ and $\lenk$ (cf. figure \ref{fig:uuscaling}b,d,f).
Similarly, the viscous diffusion term of the $\langle u'u'\rangle$ collapses very well under $\ubulk^3/h$ and $\lenvisc$ normalisation.
However, the viscous diffusion terms in $\langle v'v'\rangle$ and $\langle w'w'\rangle$ budget exhibit different scaling behaviours, where the $\langle v'v'\rangle$ diffusion term becomes similar under $\ubulk\utau^2/h$--$\lenk$ normalisation (cf. figure \ref{fig:uuscaling4}d), whereas the $\langle w'w'\rangle$ viscous diffusion term shows $\ubulk^3/h$--$\lenk$ scaling (cf. figure \ref{fig:uuscaling2}f).

The turbulent transport term in the $\langle v'v'\rangle$ budget exhibits $\ubulk\utau^2/h$--$\lenvisc$ scaling behaviour (cf. figure \ref{fig:uuscaling4}c), whilst no clear scaling behaviours were found for the turbulent transport terms in the $\langle u'u'\rangle$- and the $\langle w'w'\rangle$-budgets.
Recall that, in the near-surface region, the turbulent transport term in the TKE budget scales only partially with $\ubulk\utau^2/h$--$\lenvisc$.
The component-wise scaling analysis therefore indicates that the partial collapse of the TKE turbulent-transport term reflects heterogeneous scaling behaviour among the individual Reynolds-normal-stress components.

%
%

\clearpage
\section{Additional scaling of pressure-strain terms} \label{app:p-strain-additional}

\begin{figure} 
\centering
\includegraphics[width=\textwidth]{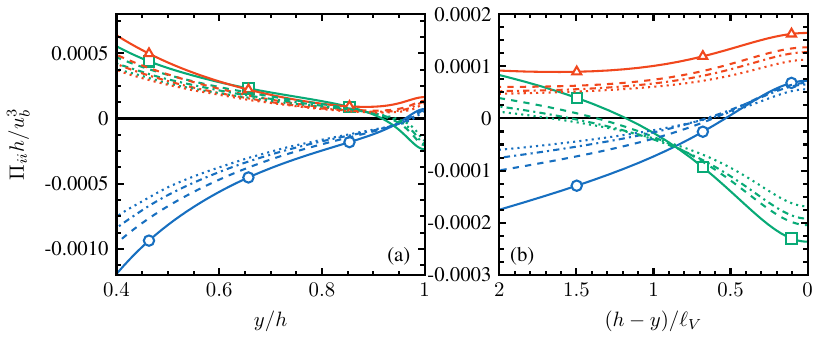}
\caption{Diagonal components of the pressure-strain term, here normalised by $\ubulk^3/h$. 
Solid (\blackline) lines, O200; dashed (\blackdashed) lines, O400; dash-dotted (\blackdashdotted) lines, O600; dotted (\blackdotted) lines, O900.
\symbx, $\Pi_{11}$; \symby, $\Pi_{22}$; \symbz, $\Pi_{33}$. 
The distance from the free surface is normalised with (a) the channel height $h$; (b) the near-surface viscous length scale $\lenvisc$.}
\label{fig:pstrainscaling2}
\end{figure}

\begin{figure} 
\centering
\includegraphics[width=\textwidth]{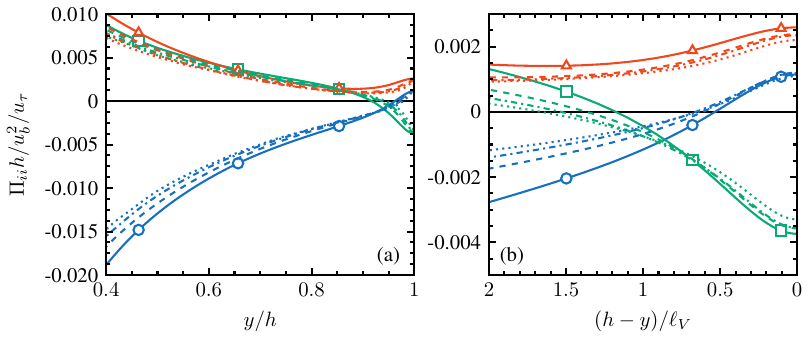}
\caption{Diagonal components of the pressure-strain term, here normalised by $\ubulk^2 \utau/h$. 
Solid (\blackline) lines, O200; dashed (\blackdashed) lines, O400; dash-dotted (\blackdashdotted) lines, O600; dotted (\blackdotted) lines, O900.
\symbx, $\Pi_{11}$; \symby, $\Pi_{22}$; \symbz, $\Pi_{33}$. 
The distance from the free surface is normalised with (a) the channel height $h$; (b) the near-surface viscous length scale $\lenvisc$.}
\label{fig:pstrainscaling3}
\end{figure}

Here, we present the bulk scaling (figure \ref{fig:pstrainscaling2}) and an additional mixed velocity scaling (figure \ref{fig:pstrainscaling3}) for the pressure-strain terms, for completeness.

\end{appendix}
\clearpage
\bibliographystyle{jfm}
\bibliography{refs_used}
\end{document}